\documentclass[11pt]{article}
\usepackage{amsmath,amsbsy,amsfonts,amssymb,graphics,epsfig,float,mathrsfs,amsthm,braket}
\usepackage{graphicx}
\usepackage{enumerate}
\usepackage{derivative}
\usepackage[normalem]{ulem}

\usepackage{graphicx}% Include figure files
\usepackage{bm}% bold math
\usepackage{enumerate} % do not remove, needed for roman numerals 
\usepackage{derivative}
\usepackage{mathtools}
\usepackage{xparse}

\newcommand{\DP}{\Delta\hspace{-0.06cm}P}
\newcommand{\wand}{\quad\text{and},\quad}
\newcommand{\wfor}{\quad\text{for}\quad}
\usepackage[colorlinks=true, allcolors=blue]{hyperref}
\newcommand{\multia}[2][]{\begin{subequations}\label{#1}
    \begin{align}
        #2
    \end{align}
\end{subequations}}
\newcommand{\multig}[2][]{\begin{subequations}\label{#1}
    \begin{gather}
        #2
    \end{gather}
\end{subequations}}
\DeclareDocumentCommand{\A}{ m m g }{
  A_{#1}^{(#2)}%
  \IfValueT{#3}{(#3)}%
}
\DeclareDocumentCommand{\St}{ m m g }{
  S_{#1}^{(#2)}%
  \IfValueT{#3}{(#3)}%
}
\DeclareDocumentCommand{\Sdt}{ m m g }{
  \mathcal{S}_{#1}^{(#2)}%
  \IfValueT{#3}{(#3)}%
}
\DeclareDocumentCommand{\Ad}{ m m g }{
  \mathcal{A}_{#1}^{(#2)}%
  \IfValueT{#3}{(#3)}%
}

\newcommand{\dA}[4]{\odv[order=#1]{A_{#2}^{(#3)}}{#4}}

\newcommand{\N}[1]{\mathcal{B}_{#1}}
\newcommand{\Nt}[1]{\mathcal{T}_{#1}}

\renewcommand{\vec}[1]{\boldsymbol{#1}}
\newcommand{\e}{\mathrm{e}}
\newcommand{\td}[2]{\frac{\mathrm{d}#1}{\mathrm{d}#2}}
\newcommand{\tdd}[2]{\frac{\mathrm{d}^2#1}{\mathrm{d}#2^2}}

\usepackage[colorlinks=true, allcolors=blue]{hyperref}

\usepackage{graphicx}% Include figure files
\usepackage{dcolumn}% Align table columns on decimal point
\usepackage{bm}% bold math

\renewcommand{\sout}[1]{}
\renewcommand{\vec}[1]{\boldsymbol{#1}}
\usepackage{ifsym}
\usepackage{psfrag}
\usepackage{color}
\usepackage{caption}
\usepackage{subcaption}
\usepackage{multirow}
\usepackage{float}
\usepackage[normalem]{ulem}

\newcommand \beq{\begin{equation}}
\newcommand \eeq{\end{equation}}

\newcommand{\LR}[1]{\left(#1\right)}

\begin{document}
\title{Snap-through time of arches is controlled by slenderness and
imperfections}

% \author{\textsf{William Simpkins$^{1}$, Matthew G.~Hennessy$^{1}$ and Matteo Taffetani$^{2}$}\\ 
% {\it$^{1}$School of Engineering Mathematics and Technology, \\ \it University of Bristol, Bristol BS8 1TW, United Kingdom}\\
% {\it$^{2}$ Institute for Infrastructure and Environment, \\ \it The University of Edinburgh, Edinburgh EH9 3FG, United Kingdom}}
\author{
\textsf{William Simpkins$^{1}$, Matthew G.~Hennessy$^{1}$ and Matteo Taffetani$^{2}$}\\
\textit{$^{1}$School of Engineering Mathematics and Technology,}\\
\textit{ University of Bristol, Bristol BS8 1TW, United Kingdom}\\
\textit{$^{2}$Institute for Infrastructure and Environment,}\\
\textit{ The University of Edinburgh, Edinburgh EH9 3FG, United Kingdom}
}

\date{}
\maketitle
\hrule\vskip 6pt
\begin{abstract}
Snap-through occurs in elastic structures when a stable equilibrium configuration becomes unstable, resulting in rapid motion towards a new and distinct stable state.  While static analyses of snap-through are well documented, the dynamics of snap-through remain under-explored, particularly in structures with natural curvature.  Using a combination of finite element simulations and multiple-scales analysis, we show that the snap-through dynamics of an arch under a central point load are controlled by its slenderness and imperfections embedded in the system.  As the slenderness increases, the
snap-through dynamics slow down, and the mode of snap-through changes
from limit-point buckling to bifurcation buckling.  When bifurcation buckling occurs, snap-through is preceded by an extended period of oscillatory behaviour. The duration of these pre-snap-through oscillations, and hence the snap-through time, is entirely controlled by imperfections in the system.  Increasing the strength of imperfections dramatically reduces the snap-through time. Analytical expressions for the snap-through times are presented for limit-point and bifurcation buckling. Our work suggests that natural curvature and deliberately introduced imperfections can be used to  tune the snap-through dynamics of new functional materials.
\end{abstract}
\vskip 6pt
\hrule

%\maketitle

\vspace{1cm}
Many biological and technological structures achieve functionality by moving between two
distinct equilibrium configurations via
snap-through instabilities
as a load or stimulus is varied~\cite{Cusp}.
These snap-through instabilities occur when the original equilibrium
configuration becomes unstable, at which
point stored elastic energy is converted into
kinetic energy, causing rapid deformation
of the structure towards a new configuration.  
Snap-through instabilities allow plants and animals to achieve fast motions
that would be impossible using muscular 
action~\cite{Venus, Hummingbird} and have been
harnessed to induce ultra-fast motion in polymer gels~\cite{JumpingPolymer},
improve the response time of soft robots~\cite{li2022, Soft-Robotics-Propultion}, enhance the controllability of smart materials~\cite{valves}, and carry out logic
operations~\cite{Logic, Thermal}.  

New metamaterials for energy harvesting \cite{Energy-Harvesting, Trapping-Elastic-Energy}, deployable structures \cite{deployable-structures}, and mechanical computation \cite{Logic} have been created 
by chaining together snap-through events and making use of elastic
hysteresis~\cite{Bi-Stable-Lattice}, multi-stability~\cite{Tristabillity}, and
tailored imperfections \cite{Guided-Transition-Waves}.  The response time of sequentially
buckling metamaterials is set by the speed of transition waves that propagate
through the structure, which is linked to the snap-through time of individual elements.
The snap-through time, defined as the time needed to snap between two equilibrium
configurations, also limits the actuation rate of materials driven by 
electrical \cite{Electrical}, magnetic \cite{magnetic}, and optical \cite{optical} stimuli.
Thus, a comprehensive understanding of the factors affecting the snap-through time is needed to fully exploit the enhanced functionalities of multi-stable structures.

Historically, the study of snap-through
instabilities has focused on computing
and characterising equilibria, e.g.\ through
the energy landscape~\cite{Cusp,thompson,MagneticandMechanical}, with the aim of
tuning the onset of instability. Analyses based on nonlinear
dynamics have revealed that two modes of snap-through can occur, termed
``limit-point'' and ``bifurcation'' buckling, originating from limit-point
and pitchfork bifurcations in the system, respectively~\cite{thompson}. 
Recent interests have shifted towards understanding the dynamic
transitions between stable equilibria~\cite{Gomez2016, Radisson_2023,Wang}.

Gomez et al.~\cite{Gomez2016} 
showed that the delayed dynamics of snap-through of a compressed
beam with asymmetric boundary 
rotation is an elastic, rather than viscoelastic~\cite{Venus,Pseudo-bistable}, bifurcation-induced phenomenon.
By examining the same system but
varying the symmetries in the boundary
conditions, Radisson and Kanso \cite{Radisson_2023,Radisson_symmetries} showed that a compressed beam can 
undergo both limit-point and bifurcation
buckling, the latter of which is distinguished by an initial period
where the beam oscillates before exponentially snapping through to its new
configuration.  Radisson and Kanso proposed a scaling law for the snap-through
time associated with bifurcation buckling; however, their calculations of specific snap-through times required an ad-hoc initial velocity, obtained from fitting to numerical simulations, suggesting the snap-through dynamics have yet to be fully
elucidated.

In this letter, we ask the question: how
does the natural curvature of an arch impact
its snap-through dynamics?  The answer is surprisingly
subtle and requires a complete understanding of the dynamics of bifurcation buckling, which we develop
through a systematic multiple-scales analysis. 
The analysis reveals the non-trivial, yet critical, role that 
intrinsic and imposed imperfections play in bifurcation buckling.

To this end, we consider a slender and shallow
arch with natural curvature $1/R^*$, 
cross-sectional area $A^*$, second moment of
area $I_x^*$, 
opening angle $2 \Theta$, density $\rho^*$, Young's modulus $E^*$, and
sound speed $c^* = \sqrt{E^*/\rho^*}$, whose schematic is provided in the SM (see Supplementary Material Sec.~\ref{Equations of motion}).
Stars denote dimensional quantities.
The arch is acted on by a central load, which can trigger snap-through~\cite{pi2002}.  

The relevant dimensionless parameters are the slenderness $\lambda = R^*\Theta^2\sqrt{A^*/I^*_x}$, which provides a measure of the arch curvature, and the load intensity $P$.
By introducing the elastodynamic time scale
$t_e^* = R^* / c^*$ and radial and circumferential
displacement scales $R^* \Theta^2$ and
$R^* \Theta^3$, respectively, the non-dimensional equations for the arch are~\cite{Pi2012} (see Supplementary Material Sec.~\ref{Equations of motion})

\begin{subequations}
\label{PDEsLetter}
\begin{gather}
    \pdv[order=2]{v}{t}+\frac{1}{\lambda^2}\pdv[order=4]{v}{\theta}-\pdv{}{\theta}\left(\varepsilon_m  \pdv{v}{\theta}\right)-\varepsilon_m=\frac{P}{\lambda^2}\delta(\theta)\label{radial PDE Letter},\\
    \pdv[order=2]{w}{t}=\pdv{\varepsilon_m}{\theta},
\end{gather}
\end{subequations}
where $\theta = \theta^*/\Theta$ is the rescaled circumferential angle, $t$ is time, $\delta(\cdot)$ is the Dirac delta function and $\boldsymbol{u} = (v,w)$ is the displacement field with radial ($v$) and circumferential ($w$) components. The membrane strain is $\varepsilon_m = \partial_\theta w-v+(1/2)(\partial_\theta v)^2$. 
Hinged boundary conditions are applied.

\begin{figure}
    \centering
    \includegraphics[width=0.8\linewidth]{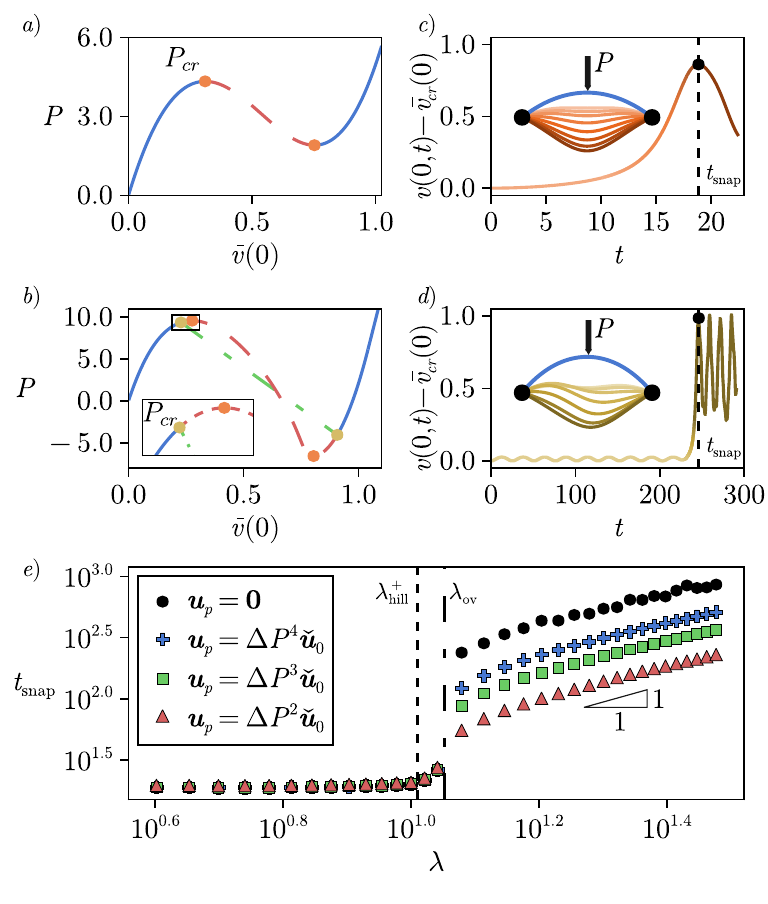}
    \caption{Equilibrium load $P$ vs midpoint displacement $\bar{v}(0)$ curves for (a) limit-point $(\lambda = 6)$ and  (b) bifurcation $(\lambda=12)$ buckling.  Solid, dashed, and dash-dot lines represent stable symmetric, unstable symmetric, and unstable asymmetric solutions, respectively. Arch midpoint displacements during (c) limit-point and (d) bifurcation buckling.  The insets show the arch shape with colour gradients denoting the passage of time; the initial shape is blue. (e) Snap-through time as a function of the slenderness $\lambda$ for ideal initial conditions
    (circles) and imperfect initial conditions (crosses, squares and triangles) with $\DP=10^{-2}$.    
    Transitions between buckling dynamics occur at $\lambda^+_\text{hill}$ and $\lambda_\text{ov}$.}
    \label{fig:snap displacements and time}
\end{figure}
By increasing the slenderness $\lambda$, the arch system under point indentation encompasses three regimes \cite{Bradford2002}     (see Supplementary Material Sec.~\ref{Bifurcation analysis}): 
(i) for $0 < \lambda < \lambda_c \simeq 3.9$, the arch is monostable; 
(ii) for $\lambda_c < \lambda < \lambda_\text{hill}^{+} \simeq 10$, the arch is bistable and undergoes limit-point
buckling (Fig.~\ref{fig:snap displacements and time}~(a));
(iii) for $\lambda_\text{hill}^{+} < \lambda$, the arch is bistable and undergoes
bifurcation buckling (Fig.~\ref{fig:snap displacements and time}~(b)).  
When $\lambda = \lambda_\text{hill}^{+}$, the limit-point and pitchfork
bifurcations occur at the same critical load, resulting in a hill-top bifurcation. A linear stability
analysis shows (see Supplementary Material Sec.~\ref{sec:lsa}) that the
system has intrinsic symmetries that impact
the shape of the arch during snap-through.
In the case of limit-point buckling, the
critical buckling eigenmode $\check{\vec{u}}_0$ with
zero eigenvalue
is symmetric 
about $\theta = 0$
and satisfies $(\check{v}_{0}^S(-\theta),\check{w}_{0}^S(-\theta))= (\check{v}_{0}^S(\theta),-\check{w}_{0}^S(\theta))$. For
bifurcation buckling, the critical mode is
antisymmetric and satisfies $(\check{v}_{0}^A(-\theta),\check{w}_{0}^A(-\theta))= (-\check{v}_{0}^A(\theta),\check{w}_{0}^A(\theta))$.  
In both cases, the non-critical eigenmodes
can be separated into symmetric modes
$\{\check{\vec{u}}_n^S\}$ and antisymmetric modes
$\{\check{\vec{u}}_n^A\}$ with $n \geq 1$.  The growth rates of the non-critical modes are
imaginary, implying their linear response to the point load is oscillatory in time.

To study the nonlinear dynamics of snap-through, we carry out numerical simulations (see Supplementary Material Sec.~\ref{sec:Numerics}) of eqs.~\eqref{PDEsLetter}. We first quasi-statically reach the bifurcation point by incrementing the
applied load $P$.  The displacement and
load at the buckling threshold are denoted
by $\bar{\boldsymbol{u}}_{cr} = (\bar{v}_{cr},\bar{w}_{cr})$ and $P=P_{cr}(\lambda)$, where $\bar{\vec{u}}_{cr}$ is symmetric.  We then perturb
the system by increasing the load by a small
amount $\Delta P = (P - P_{cr}) / P_{cr} \ll 1$,
forcing the system to dynamically leave the critical configuration, and
solve eqs.~\eqref{PDEsLetter} with the initial conditions
\begin{equation}\label{eq:IC_ideal}
\vec{u}(\theta,0) = \bar{\vec{u}}_{cr},
\quad
    \pdv{\boldsymbol{u}}{t}(\theta,0)=\boldsymbol{0}.
\end{equation}
During limit-point buckling, the midpoint of 
the arch monotonically increases until the
everted configuration is reached
and the arch remains symmetric (Fig.~\ref{fig:snap displacements and time}~(c)).  However, during bifurcation buckling, there is
an extended period where the arch midpoint
oscillates about the symmetric, now unstable, equilibrium before
snapping through (Fig.~\ref{fig:snap displacements and time}~(d)) (see Supplementary Material Sec.~\ref{Pre-Snap-Through Oscillations}).  
These
small-scale oscillations are analogous to
the pre-snap-through oscillations observed
by Radisson and Kanso \cite{Radisson_2023}.
Numerical experiments show that the oscillation frequency is
independent of $\Delta P$
but strongly dependent on $\lambda$ (see Supplementary Material Sec.~\ref{Pre-Snap-Through Oscillations}). As
$\lambda \to \lambda_\text{hill}^{+}$, the frequency tends to zero, 
showcasing a new type of critical slowing down that occurs when
the system approaches the hill-top bifurcation. 
The arch remains symmetric during the 
oscillatory
phase, becomes asymmetric during snap
through, and returns to being symmetric
as the new equilibrium configuration is approached (inset Fig.~\ref{fig:snap displacements and time}~(d)).  By computing the snap-through time $t_{\text{snap}}$ across a range of
slenderness values $\lambda$, we find that the rate of snap-through is vastly different for the
two buckling modes.
As shown in Fig.~\ref{fig:snap displacements and time}~(e), $t_{\text{snap}}$ is approximately independent of $\lambda$ in the
regime of limit-point buckling, and it becomes linearly dependent on $\lambda$ in the
regime of bifurcation buckling.

Unexpectedly, the snap-through
times for bifurcation buckling 
have a weak noise-like variability
that cannot be removed by refining the mesh,
time step, or numerical method; see the black circles in Fig.~\ref{fig:snap displacements and time}~(e) for $\lambda > \lambda_\text{hill}^+$.  
However, running simulations with a small imperfection introduced through 
the initial conditions,
\begin{equation}\label{eq:IC_imperfection}
     \boldsymbol{u}(\theta,0) = \bar{\vec{u}}_{cr}(\theta)+\DP\boldsymbol{u}_{p}(\theta),
     \quad
     \pdv{\boldsymbol{u}}{t}(\theta,0)=\boldsymbol{0},
\end{equation}
where $\boldsymbol{u}_{p}$ is the arbitrary shape of the imperfection, eliminates the variability in the snap-through times;
see the crosses, squares, and triangles in Fig.~\ref{fig:snap displacements and time}~(e).
Moreover, 
increasing the magnitude of the imperfection
decreases the snap-through time.  
The snap-through times associated with the limit-point buckling are, instead, unaffected by the imperfection. The simulations are carried out by setting the imperfection to $\boldsymbol{u}_{p}=\Delta P^\alpha\check{\boldsymbol{u}}_{0}$ where $\check{\vec{u}}_0$ is
the critical buckling mode (see Supplementary Material Sec.~\ref{sec:criticalmodes})
and $\alpha = 1, 2, 3$.

The dependence of the snap-through time on $\lambda$ can be estimated from a scaling
analysis. Since snap-through brings a curved arch close to its everted configuration, it is natural to scale the radial displacement as $v^* \sim R^*\Theta^2$. Following geometrical arguments~\cite{Matteo,Pauchard}, the membrane strain scales as $\varepsilon^*_m \sim v^*/\sqrt{R^* v^*}$, so that the stretching energy density is $\mathcal{E}^*_S\sim E^* (\varepsilon^*_m)^2 \sim E^* \Theta^2$. We expect the membrane strain  to be independent of the system's slenderness, since stretching strain is related to changes in the length of the system's centre line \cite{benedettini2012}. The strain associated with the change in curvature is $\varepsilon^*_b \sim h^*/R^*$, where
$h^* \sim (I^*_x/A^*)^{1/2}$ is the arch thickness. The bending energy density can be estimated as $\mathcal{E}^*_B \sim E^* (\varepsilon^*_b)^2\sim E^*\Theta^4/\lambda^2$. 
The kinetic energy associated with snap-through is given by $\mathcal{E}^*_K\sim\rho^* (v^* / t_\text{snap}^*)^2$.  
For $\lambda$ large, we balance kinetic and bending energies to obtain $t_\text{snap}^*\sim \lambda t^*_e$. 
For smaller values of $\lambda$, we
balance kinetic and stretching energy to find $t_\text{snap}^*\sim t^*_e$. 
While these arguments explain the empirical 
scaling laws found from numerical
simulations ({Fig.~\ref{fig:snap displacements and time}~(e)}),
they do not capture the critical role that
imperfections play in bifurcation buckling
nor the slowing down that occurs as
$\Delta P$ tends to zero, in both the snap-through regimes~\cite{Gomez2016}.

To resolve the dynamics of bifurcation
buckling, 
in which the arch oscillates on
a fast $O(1)$ time scale and snaps through on a slow, yet to be determined, $O(\tau)$ time scale,  we carry out a multiple-scales analysis~\cite{kevorkian1996multiple} in the limit $\Delta P \to 0$.  We
consider general imperfections in the
initial condition \eqref{eq:IC_imperfection} that are written
in terms of symmetric and
antisymmetric eigenmodes,
\begin{align}
\vec{u}_p(\theta) = p_0^A \check{\vec{u}}^A_0(\theta)
+ \sum_{n=1}^{\infty} \left[p_n^A \check{\vec{u}}^A_n(\theta) + p_n^S \check{\vec{u}}_n^S(\theta)\right],
\end{align}
where the $p_n^A$ and $p_n^S$ are
Fourier-like coefficients characterising the
amplitude of the imperfection along 
the eigenmodes.
The applied load is written as $P = P_{cr}(1 + \Delta P)$. 
In contrast to past works~\cite{Radisson_2023}, which only resolve
the dynamics of bifurcation buckling on the slow $O(\tau)$ time 
scale, 
our multiple-scales analysis simultaneously
captures the 
$O(1)$ and $O(\tau)$ time scales, allowing the 
impact of fast
oscillations on slow snap-through to be
determined.  Moreover, our approach allows
the slow time scale $\tau$ to be rigorously derived without relying on ad-hoc arguments or numerical considerations.

We find that oscillations play an
essential role in triggering bifurcation
buckling, which occurs across three distinct time regimes given by
$t = O(1)$, $t = O(\Delta P^{-1/2})$, and $t = O(\Delta P^{-1/2}\ln(1/\Delta P))$. A detailed
analysis of these regimes can be found
in the supplementary material (see Supplementary Material Sec.~\ref{sec:BifBuck}).
When $t = O(1)$, nonlinear
interactions between oscillatory symmetric modes and the antisymmetric
critical mode lead to resonance, causing a monotonic growth of the critical mode.  The amplitude of the critical mode remains smaller than the amplitudes of the symmetric
modes.  The amplitude of each symmetric mode is inversely proportional
to the square of its natural frequency (in time).  Thus,
the behaviour of the arch is dominated by the lowest-frequency
symmetric mode, leading to the emergence of pre-snap-through
oscillations.
When $t = O(\Delta P^{-1/2})$, the
sustained resonant interactions between
oscillatory symmetric modes and the critical
mode initiate snap-through.  
However, the amplitude of the critical mode remains small and its motion is governed by a linear differential equation, resulting in exponential growth.
Finally, when $t = O(\Delta P^{-1/2}\ln(1/\Delta P))$, the exponential growth of the critical mode initiates nonlinear effects.  Due to the growth of
the critical mode, its amplitude becomes much greater than all other 
modes.  The arch displacement is then approximately given by the
superposition of the symmetric
equilibrium displacement $\bar{\vec{u}}_{cr}$
and the antisymmetric critical mode $\check{\vec{u}}_0$, resulting in an asymmetric shape
during snap-through.

In the final time regime of bifurcation buckling,
the equation of motion for the amplitude of the critical mode is the normal form of a
pitchfork bifurcation.  The solutions of
this nonlinear differential equation exhibit finite-time blow up.  We associate the blow-up time with the snap-through time of the arch~\cite{Gomez2016}.  By solving the
differential equation and matching to the
exponentially small solution from the
$O(\Delta P^{-1/2})$ time regime, we 
obtain a novel analytical expression for
the snap-through time for bifurcation
buckling,
\begin{equation}\label{Asymmetric snapthrough time}
    t^\text{B}_\text{snap} = \frac{{\lambda}}{\sqrt{r_\text{B}\DP}}\ln\left(\frac{4}{p^A_0}\sqrt{\frac{2r_\text{B}}{q_\text{B}\DP}}\right),
\end{equation}
where $r_\text{B}$ and $q_\text{B}$ are functions of $\lambda$ that can be computed analytically (see Supplementary Material Sec.~\ref{sec:BifBuck2}).  As $\lambda \to \infty$, both $r_\text{B}$ and $q_\text{B}$ tend to constants, so $t_\text{snap}^\text{B} = O(\lambda$),
consistent with the scaling analysis and simulations. 
The appearance
of the coefficient $p_0^A$ in \eqref{Asymmetric snapthrough time} 
quantifies how imperfections impact the
snap-through time.  As $p_0^A$
decreases, the snap-through time 
logarithmically increases. Importantly, the
snap-through time is only influenced by
imperfections that have a component along
the critical buckling mode, the magnitude of which is encoded in
$p_0^A$.

The central point load
leads to a symmetric actuation that only
excites the symmetric eigenmodes of the 
system.  The symmetric pre-snap-through oscillations are thus the leading-order response of the system on $O(1)$ time scales.
The critical mode, which is antisymmetric in
bifurcation buckling,
can only be excited through nonlinear interactions between itself
and the symmetric modes.  
Therefore, if there are no imperfections, or if
the imperfection does not have a component
along the critical
mode, e.g.\ the imperfection is
symmetric, then the critical mode cannot be 
excited and it will never grow.  
In this case, the oscillations persist
indefinitely and snap-through does not
occur, despite the threshold of bifurcation
buckling being exceeded.  Indeed, taking
$p_0^A \to 0$ in \eqref{Asymmetric snapthrough time} shows that the snap-through
time blows up, $t_\text{snap}^\text{B} \to \infty$.
When the oscillations persist,
it can be shown that the oscillation period
undergoes a slow modulation on an 
$O(\Delta P^{-1})$ time scale.
This is much greater than the time scale of snap-through, which is
why we find that the period of the oscillations does not depend on
$\Delta P$ in the numerical simulations.

Imperfection is what sets the finite snap-through time in the case of bifurcation buckling: increasing the size of the imperfection accelerates snap-through, consistent with the simulation results in Fig.~\ref{fig:snap displacements and time}~(e). The \emph{noisy} behaviour that we highlighted earlier
can thus be explained by the numerical discretization of the problem introducing small imperfections, but whose specific value cannot be controlled systematically for all values of the
slenderness $\lambda$.

The interplay between imperfections and oscillations in bifurcation buckling has been discussed in the context of compressed beams with time-dependent loading, which exhibit precursor oscillations before
the bifurcation threshold is 
crossed~\cite{Wang, Imperfections}. 
The loading rate was shown to control the nature of snap-through and 
determine how imperfections are amplified
through precursor oscillations to produce
transient asymmetries.
Our system is different in that the loading is static, and static loading causes the arch to oscillate about a globally unstable equilibrium, rather than
dynamic loading causing oscillations about a stable equilibrium.
In our case, imperfections are a necessary precondition for, and completely dictate the speed of, bifurcation buckling.

\begin{figure}
    \centering
    \includegraphics[width=0.8\linewidth]{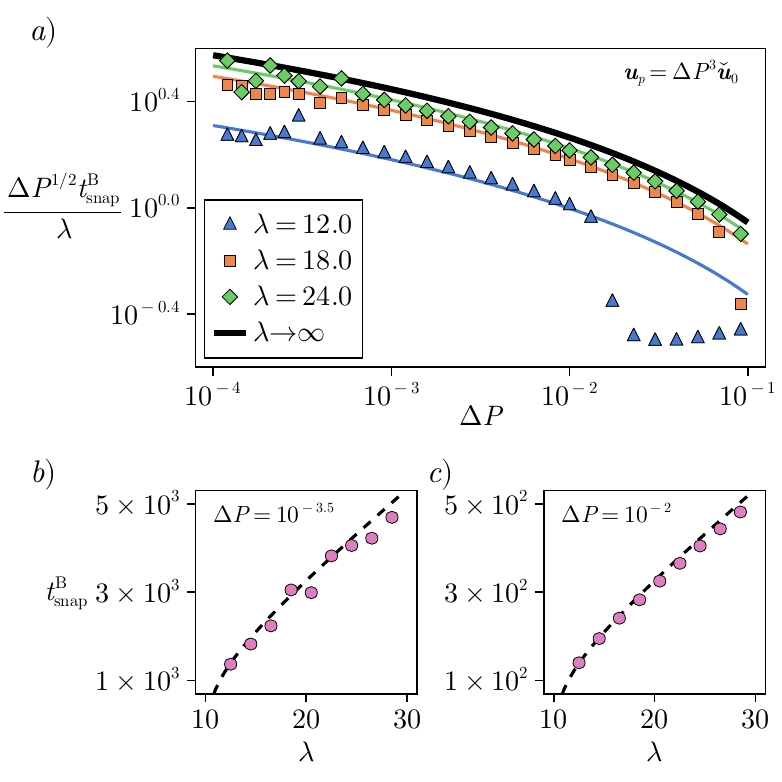}
    \caption{(a) Rescaled snap-through times 
    for bifurcation buckling
    as a function of
    $\DP$ for different arch slendernesses $\lambda$ and an applied imperfection of size $\DP^3$.     
    Markers indicate times 
    from finite-element simulations. Snap-through time as a function of $\lambda$ for an applied imperfection that is (b) smaller and (c) greater than intrinsic imperfections. Solid and dashed curves are theoretical predictions.}
    \label{Snap times accross DP plot}
\end{figure}

Imperfections do not only fix the snap-through time, but also affect the slowing down induced by the pitchfork bifurcation: the logarithmic contribution to the snap-through time in \eqref{Asymmetric snapthrough time} represents an additional slowing down that has not been predicted in previous analyses of
bifurcation buckling~\cite{Radisson_2023}. Plotting the rescaled snap-through time
$\Delta P^{1/2} \lambda^{-1} t^\text{B}_\text{snap}$
computed from numerical simulations as a function of $\Delta P$ shows that the data do not lie on a horizontal line (Fig.~\ref{Snap times accross DP plot}~(a)), highlighting the need to account for the logarithmic factor.  Indeed, the analytical predictions from \eqref{Asymmetric snapthrough time}, shown as solid curves in
Fig.~\ref{Snap times accross DP plot}~(a) for selected values of $\lambda$, accurately capture the numerically computed 
snap-through times. 
Applied imperfections can only control the snap-through time if their effect is larger than intrinsic imperfections of the system, which in our case arise from numerical discretisation.
For small values of $\Delta P$,
the applied imperfection, which has amplitude $p_0^A = \Delta P^3$, becomes smaller than intrinsic imperfections, and the numerical snap-through times in Fig.~\ref{Snap times accross DP plot}~(b) show the same \emph{noisy} behaviour as when no imperfections are applied.  For larger values of $\Delta P$, the
applied imperfection is greater than the intrinsic imperfections, and the numerical snap-through times vary smoothly with controllable system parameters (Fig.~\ref{Snap times accross DP plot}~(c)).

For large $\lambda$, the numerical data in Fig.~\ref{Snap times accross DP plot}~(a) approach a universal curve, consistent with the prediction that 
$t_\text{snap}^\text{B} \sim \lambda$.  
The universal behaviour is not followed when the slenderness is close to the threshold $\lambda_\text{hill}^+ \simeq 10$.  
In this case, the
limit-point and pitchfork bifurcations
occur for similar critical loads. Increasing the applied load by
$\Delta P$
can result in an ``overshoot'', whereby limit-point buckling is triggered alongside bifurcation buckling (see Supplementary Material Sec.~\ref{Sec:Overshoot}), 
reducing the snap-through time.  Alternatively, for a fixed
$\Delta P$, there is a range of slenderness values $\lambda_\text{hill}^{+} < \lambda < \lambda_\text{ov}(\Delta P)$
for which overshoot occurs.  The critical value 
$\lambda_{\text{ov}}$, rather than $\lambda_\text{hill}^{+}$, determines whether limit-point
or bifurcation buckling occurs and hence controls the
snap-through time (Fig.~\ref{fig:snap displacements and time}~(e)).

The dynamics of bifurcation buckling do not depend on the
nature of the imperfection.  By replacing the imperfection in the initial
condition \eqref{eq:IC_imperfection} with a geometric imperfection, the same hierarchy of time regimes
emerges (see Supplementary Material Sec.~\ref{sec:geo_imperfections}).  The snap-through time is given by \eqref{Asymmetric snapthrough time}, but the
parameter $p_0^A$ now depends on the perturbation to the critical buckling
configuration $\vec{u}_{cr}$ induced by the imperfection.

There are no pre-snap-through oscillations
during limit-point buckling. 
The multiple-scales analysis (see Supplementary Material Sec.~\ref{sec:LP}) shows that snap-through occurs across two time regimes.  When
$t = O(1)$, the applied load excites all 
of the symmetric modes, including the critical mode.  The amplitude of the critical
mode grows quadratically in time due to the load and quickly
overtakes the oscillatory amplitudes of the non-critical modes, thus dominating the arch behaviour.
When $t = O(\Delta P^{-1/4})$, the amplitude of the critical 
mode is influenced by nonlinear interactions with itself, and
its equation of motion is the normal form of
a limit-point (or saddle-node) bifurcation.  Solving this
equation furnishes a snap-through time of
\begin{equation} \label{Symmetric snapthrough time}
    t_\text{snap}^\text{LP} = \left(\frac{3}{64\pi^2r_\text{LP}q_\text{LP}\DP}\right)^\frac{1}{4}\Gamma\left(\frac{1}{4}\right)^2,
\end{equation}
where $\Gamma(\cdot)$ is the gamma function.
We recover the slowing down $t_\text{snap}^\text{LP} \sim \DP^{-1/4}$ induced by the limit-point bifurcation~\cite{Gomez2016, Radisson_2023}, but we also capture the role of the slenderness through the functions $r_{\text{LP}}\left(\lambda\right)$ and $q_{\text{LP}}\left(\lambda\right)$ (see Supplementary Material Sec.~\ref{sec:LP}).  Although $r_{\text{LP}}$ and $q_{\text{LP}}$ depend on $\lambda$, the fourth root of their product remains approximately constant.  Thus, the
snap-through time \eqref{Symmetric snapthrough time} is approximately independent of $\lambda$,
consistent with the numerical results in Fig.~\ref{fig:snap displacements and time}~(e) and the
scaling analysis.  Importantly, the snap-through evolution during limit-point buckling is insensitive to applied imperfections.

Our analysis quantitatively characterizes the snap-through dynamics of an elastic arch under point indentation, revealing the role of slenderness as an internal parameter that discriminates symmetric from asymmetric evolution and controls snap-through time.
For limit-point buckling, snap-through time is robust to both slenderness and applied imperfection size. In bifurcation buckling, increasing slenderness slows evolution and requires an asymmetric imperfection to produce a finite snap-through time, with the asymmetric mode growing via resonance with preceding symmetric oscillations. The previously proposed~\cite{Radisson_2023} scaling, $t_\text{snap}^\text{B} \sim \DP^{-1/2}$, must be amended with a logarithmic correction.  
These results extend beyond arches to any mechanical system with an equivalent bifurcation landscape.

Buckling is exploited in natural and artificial systems to enhance functionality. At small scales (e.g., in micro-electronics \cite{Das2009} or fast actuators \cite{Xia2010, Zhang2022}) designing both stable equilibria and rapid dynamic transitions are equally important. Our framework offers a design tool to estimate morphing rates in slender systems where snap-through time is rate-limiting and opens the
doors to using deliberately introduced imperfections to precisely
tune the snap-through times of new functional materials.\\

M.T. is a member of the Gruppo Nazionale di Fisica Matematica (GNFM) of the Istituto Nazionale di Alta Matematica (INdAM). M.H. was supported by the Engineering and Physical Sciences Research Council [grant number UKRI093]. We thank Andrea Giudici and Dominic Vella for insightful discussions about this work.

\vskip6pt

\vskip2pc

\clearpage
\appendix       %%% starting appendix
\section*{Supplementary Material}\label{SM}

\section{Equations of motion}\label{Equations of motion}
\begin{figure}
    \centering
    \includegraphics[width=0.35\linewidth]{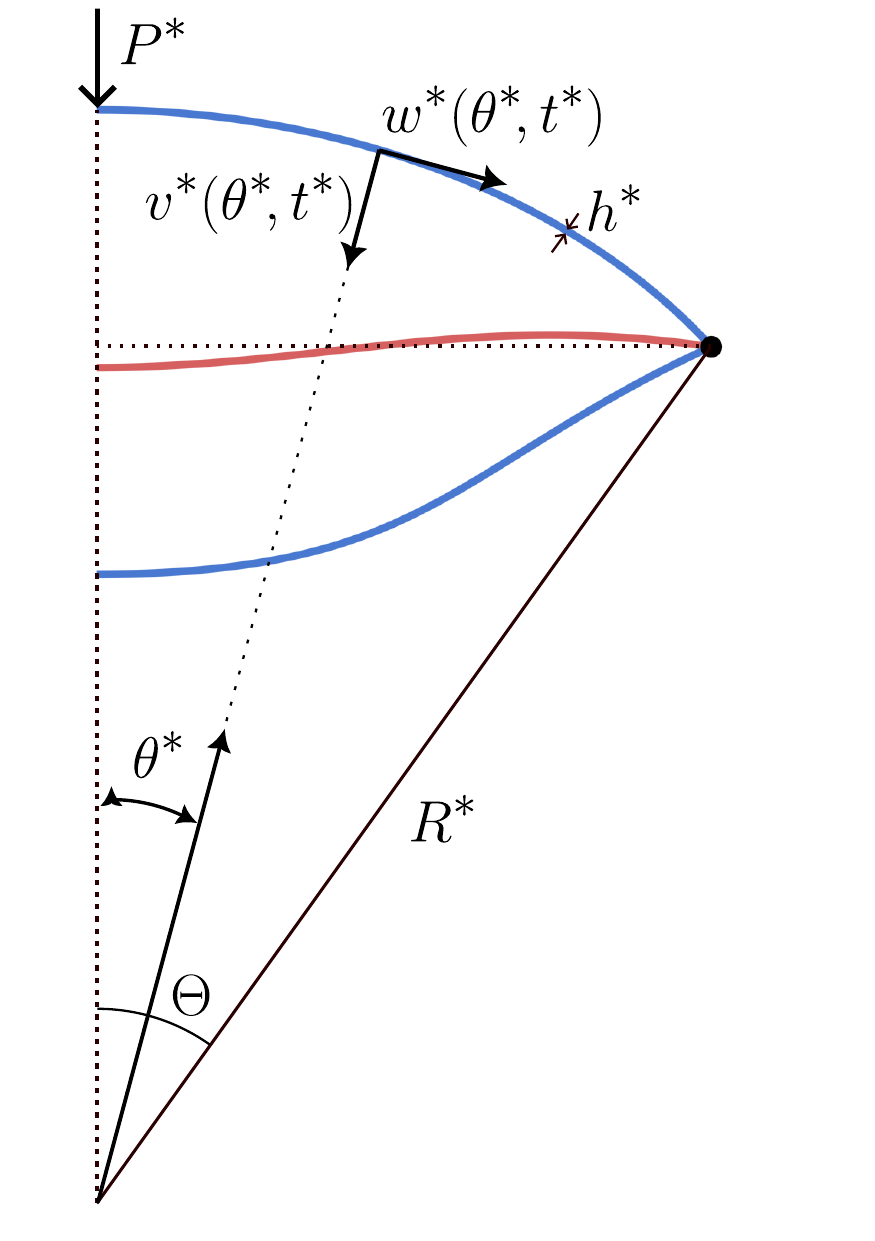}
    \caption{Sketch of an arch undergoing a snap-through instability due to a central point load.  The symbols $v^*(\theta^*,t^*)$ and $w^*(\theta^*,t^*)$ denote the radial and circumferential displacements, $P^*$ the applied load, $R^*$ the arch radius in the undeformed configuration, $\Theta$ half of the opening angle and $h^*$ the arch thickness. All terms marked with an asterisk are dimensional.}
    \label{fig:schematic}
\end{figure}

We describe the behaviour of a slender shallow arch, where lateral displacements and twist rotations are not permitted, subjected to a point load acting on its midpoint; see Fig.~\ref{fig:schematic}.
The governing equations for the radial ($v^*$) and circumferential ($w^*$) displacements are given by~\cite{Trahair}
\begin{subequations}
\begin{align}
\label{PDEs Dimensional}
    &\rho^*\pdv[order=2]{v^*}{t^*}+\frac{E^* I^*_x}{A^*(R^*)^4}\pdv[order=4]{v^*}{\theta^*}-\frac{E^*}{(R^*)^2}\pdv{}{\theta^*}\left(\varepsilon_m\pdv{v^*}{\theta^*}\right)-\frac{E^*}{R^*}\varepsilon^*_m-\frac{P^*}{R^*A^*}\delta(\theta^*)=0, \\
    &\rho^*\pdv[order=2]{w^*}{t^*}-\frac{E^*}{R^*}\pdv{\varepsilon^*_m}{\theta^*}=0,
\end{align}
\end{subequations}
where $t^*$ is time, $\theta^*$ is the circumferential angle along the centreline of the arch,
$\rho^*$ is the mass density, $E^*$ is the Young's
modulus, $R^*$ is the arch radius, $A^*$ is the cross-sectional area, $I^*_x$ is the second area moment, $P^*$ is the applied load magnitude, and $\delta$ is the Dirac delta
function. These equations account for geometrical non-linearities while keeping the constitutive relationship linear~\cite{pi2002}.
The membrane (stretching) strain is given by
 \begin{equation}
     \varepsilon^*_m = \frac{1}{R^*}\pdv{w^*}{\theta^*} - \frac{v^*}{R^*} + \frac{1}{2(R^*)^2}\left(\pdv{v^*}{\theta^*} \right)^2.
     \label{eqn:em}
 \end{equation}
We add to these equations the following hinged boundary conditions,
\begin{equation}
    v^*(\pm\Theta,t^*)=\pdv[order=2]{v^*}{\theta}(\pm\Theta,t^*)=w^*(\pm\Theta,t^*)=0,
\end{equation}
where $2 \Theta \ll 1$ is the small opening angle of the arch. 
The initial conditions are given by
\begin{equation}
    v^*(\theta^*,0) = V^*(\theta^*), \quad w^*(\theta^*,0) = W^*(\theta^*), \quad \pdv{v^*}{t}(\theta^*,0)=\pdv{w^*}{t^*}(\theta^*,0)=0,
\end{equation}
where $V^*$ and $W^*$ are initial radial and circumferential displacements. 
We will later set $V^*$ and $W^*$ to be the 
displacements associated with the
buckled state to determine the snap-through dynamics.  We will also introduce small 
imperfections into the system by adding 
perturbations to $V^*$ and $W^*$.
\subsection{Model non-dimensionalisation}
\label{sec:non-dim}
The governing equations are cast into dimensionless form by writing
\begin{align}\label{eq:nondimensionalization}
    \theta=\frac{\theta^*}{\Theta},\quad t=\frac{c^*}{R^*}\,t^*,\quad 
    v = \frac{v^*}{R^*\Theta^2},\quad 
    w = \frac{w^*}{R^*\Theta^3},\quad \varepsilon_m = \frac{\varepsilon^*_m}{\Theta^2},
    \quad
    P = \frac{(R^*)^2\Theta}{E^*I^*_x}P^*.
\end{align}
Using these expressions, the non-dimensional equations of motion are given by 
\begin{subequations}
\label{PDEs}
\begin{align}
    &\pdv[order=2]{v}{t}+\frac{1}{\lambda^2}\pdv[order=4]{v}{\theta}-\pdv{}{\theta}\left(\varepsilon_m  \pdv{v}{\theta}\right)-\varepsilon_m-\frac{P}{\lambda^2}\delta(\theta)=0\label{radial PDE},\\
    &\pdv[order=2]{w}{t}-\pdv{\varepsilon_m}{\theta}=0\label{axial PDE},
\end{align}
\end{subequations}
where $\lambda$ is the modified slenderness defined as
\begin{equation}\label{lambda}
    \lambda = R^* \Theta^2 \sqrt{\frac{A^*}{I^*_x}}.
\end{equation}
The membrane strain is given by
\begin{equation}\label{eq:membranestrain}
    \varepsilon_m = \pdv{w}{\theta} - v +\frac{1}{2}\left(\pdv{v}{\theta}\right)^2.
\end{equation}
The dimensionless boundary conditions are
\begin{equation}
    v(\pm1,t)=\pdv[order=2]{v}{\theta}(\pm1,t)=w(\pm1,t)=0.
\end{equation}
The corresponding initial conditions are
\begin{align}
v(\theta,0)= V(\theta), \quad w(\theta,0)= W(\theta), \quad \pdv{v}{t}(\theta,0)=\pdv{w}{t}(\theta,0)=0.
\label{nd:ics}
\end{align}
Integrating \eqref{eq:membranestrain} and using the boundary conditions allows us to derive a compatibility condition, relating the change in length of the mid-line of the arch to the average membrane strain field, as
\begin{equation}\label{eq:compatibilityGeneral}
    \int_{-1}^1 \varepsilon_m\, \mathrm{d}\theta  = \int_{-1}^1\left[\frac{1}{2}\left(\pdv{v}{\theta}\right)^2-v\right]\, \mathrm{d}\theta.
\end{equation}
Examining the dimensionless model given by \eqref{eq:nondimensionalization}--\eqref{eq:compatibilityGeneral} shows that the dynamics of the arch are controlled
by two dimensionless parameters: the applied load $P$ and the
modified slenderness $\lambda$.

\section{Bifurcation analysis}\label{Bifurcation analysis}

The qualitative features of the solutions to the
dimensionless model given by \eqref{PDEs}--\eqref{nd:ics}
are determined through a bifurcation analysis.  This
involves computing the equilibrium configurations
of the system and examining their linear stability.

\subsection{Equilibrium configurations}
\label{sec:equilibrium}

The equilibrium configurations of the arch are obtained by
seeking solutions that are independent of time, which are denoted with a bar.  
Equation \eqref{axial PDE} then implies that the equilibrium membrane strain $\bar{\varepsilon}_m$ is constant along the length of the arch.
Thus, we let
\begin{equation}\label{eq:strain_equilibrium}
    \bar{\varepsilon}_m=\pdv{\bar{w}}{\theta} - \bar{v} +\frac{1}{2}\left(\pdv{\bar{v}}{\theta}\right)^2 \equiv -\frac{\mu^2}{\lambda^2},
\end{equation}
where $\mu$ is a constant that is related to the force in the
circumferential direction~\cite{pi2002}.  The radial
force balance \eqref{radial PDE} reduces to
\begin{equation}\label{ODE Problem}
    \odv[order=4]{\bar{v}}{\theta}+\mu^2\odv[order=2]{\bar{v}}{\theta}+\mu^2=P\delta(\theta).
\end{equation}
There are two types of solutions to \eqref{eq:strain_equilibrium}--\eqref{ODE Problem} that are
characterised by different symmetry properties.  We define
\emph{symmetric} solutions as those which satisfy 
\begin{align}
v(-\theta, t) = v(\theta, t), 
\qquad 
w(-\theta, t) = -w(\theta, t),
\label{eqn:eq_symmetry}
\end{align}
owing to the symmetry of the radial displacement about the arch midpoint.  
The existence of symmetric solutions naturally follows from the
governing equations and can be verified through a change of
variable.  However, there are also
\emph{asymmetric} solutions that do not exhibit
any symmetry properties.  Asymmetric solutions are the result of spontaneous symmetry breaking
occurring at specific parameter combinations, as seen
by Pandey et al. \cite{Poppers} in the context of compressed
beams subject to a concentrated load.
While the symmetric solution occurs at all values of $\mu$, asymmetric solutions are admittable only for $\mu = \pi$ (for the loading condition considered here).

For a symmetric equilibrium solution, the radial displacement $\bar{v}$  can be expressed in terms of the circumferential force $\mu$ and the applied load $P$ as
\begin{equation}
    \bar{v}(\theta) = A + B |\theta| -\frac{1}{2}\theta^2+C\cos(\mu\theta)+D\sin(\mu|\theta|)\label{v static sol},
\end{equation}
where
\begin{equation}
    A = \frac{\mu^2-P+2}{2\mu^2},\quad B = \frac{P}{2\mu^2}, \quad C = \frac{P\sin(\mu)-2\mu}{2\mu^3\cos(\mu)},\quad D = -\frac{P}{2\mu^3}.
    \label{eqn:ABCD}
\end{equation}
The circumferential displacement $\bar{w}$ can be obtained by
substituting \eqref{v static sol} into \eqref{eq:strain_equilibrium} and
solving for $\bar{w}$. To determine the constant $\mu$, we use the equilibrium configuration in \eqref{eq:compatibilityGeneral} to obtain
\begin{equation}\label{static comp cond}
    F(P,\mu,\lambda) = \frac{2\mu^2}{\lambda^2}+\int_{-1}^1\left[\frac{1}{2}\left(\pdv[order=1]{\bar{v}}{\theta}\right)^2-\bar{v}\right]\, \mathrm{d}\theta = 0.
\end{equation}
Upon substitution of the equilibrium solution for $\bar{v}$ given
by \eqref{v static sol} into \eqref{static comp cond}, a
nonlinear equation that determines the circumferential force
$\mu$ in terms of the applied load $P$ and the modified
slenderness $\lambda$ arises. 

The asymmetric equilibrium solution arises when
$\mu = \pi$ and is given by
\begin{equation}
    \bar{v}(\theta) = A + B |\theta| -\frac{1}{2}\theta^2+C\cos(\mu\theta)+E\sin(\mu\theta),
\end{equation}
where 
$A$, $B$, and $C$ are given by \eqref{eqn:ABCD}
after substituting $\mu = \pi$.  The coefficient
$E$ is given by
\begin{align}
E = \begin{cases}
    K+\dfrac{P}{\pi^3},\wfor \theta<0, \\ K,\quad\quad\hspace{0.15cm}\wfor\theta\geq0,
    \end{cases}
\end{align}
where
\begin{equation}
    K = -\frac{3 \pi^{3} P \lambda +\sqrt{-144 \pi^{12}+24 \pi^{10} \lambda^{2}-27 P^{2} \pi^{6} \lambda^{2}-36 \pi^{8} \lambda^{2}+144 P \pi^{6} \lambda^{2}}}{6 \pi^{6} \lambda}.
\end{equation}

\subsection{Linear stability analysis}
\label{sec:lsa}
The stability of the equilibria computed in Sec.~\ref{sec:equilibrium} can be assessed by 
seeking a solution to the time-dependent model 
given by
\begin{equation}
    \vec{u}(\theta, t) = \bar{\vec{u}}(\theta) + \epsilon \check{\vec{u}}(\theta)\,\e^{i \omega t}, \quad \varepsilon_m = \bar{\varepsilon}_m + \epsilon \check{\varepsilon}_m(\theta)\,\e^{i \omega t},
    \label{lsa:sol}
\end{equation}
where $\vec{u} = (v, w)$ is the non-dimensional displacement
vector, $\bar{\vec{u}}$ and $\bar{\varepsilon}_m$ are the equilibrium fields, 
$\epsilon \ll 1$ is an arbitrary small parameter, and $\omega$ is a complex growth rate.
If $\omega$ is real then the system is expected
to oscillate about the equilibrium state; in this case, $\omega$
corresponds to one of the natural frequencies of the arch.
If $\omega$ is imaginary, then the perturbation to the equilibrium state will either grow or
decay, depending on the sign of the imaginary part of $\omega$.
Upon substitution of \eqref{lsa:sol} into \eqref{PDEs}--\eqref{nd:ics}, expanding about $\epsilon \ll 1$,
and neglecting small terms of $O(\epsilon^2)$, we find that the perturbations
$\check{\vec{u}} = (\check{v}, \check{w})$ 
and the growth rates $\omega$
are determined by the eigenproblem
\begin{equation}\label{first eigenproblem}
    \mathcal{L}\check{\vec{u}} = \omega^2\check{\vec{u}},
\end{equation}
where $\mathcal{L}$ is a matrix of linear differential operators given by
\begin{subequations}
\label{L operator}
\begin{align}
\mathcal{L} = \begin{pmatrix}
\mathcal{L}_{vv} & \mathcal{L}_{vw} \\
\mathcal{L}_{wv} & \mathcal{L}_{ww}
\end{pmatrix},
\end{align}
with
\begin{align}
\mathcal{L}_{vv} &= 
        \frac{1}{\lambda^2}\odv[order=4]{}{\theta}+\left(\frac{\mu^2}{\lambda^2}-\left(\odv{\bar{v}}{\theta}\right)^2\right)\odv[order=2]{}{\theta}-\odv{\bar{v}}{\theta}\left(2\odv[order=2]{\bar{v}}{\theta}+1\right)\odv{}{\theta} - \left(\odv[order=2]{\bar{v}}{\theta}+1\right), 
\\ 
\mathcal{L}_{vw} &= -\odv{\bar{v}}{\theta}\odv[order=2]{}{\theta}-\left(\odv[order=2]{\bar{v}}{\theta}+1\right)\odv{}{\theta},
\\
\mathcal{L}_{wv} &= 
        \odv{}{\theta}-\odv[order=2]{\bar{v}}{\theta}\odv{}{\theta}-\odv{\bar{v}}{\theta}\odv[order=2]{}{\theta}, 
\\
\mathcal{L}_{ww} &= -\odv[order=2]{}{\theta}.
\end{align}
\end{subequations}
The boundary conditions for \eqref{first eigenproblem} are given
by
\begin{equation}
    \check{v}(\pm 1) = \odv[order=2]{\check{v}}{\theta}(\pm 1) = \check{w}(\pm 1) = 0.
    \label{lsa:bc}
\end{equation}

It is straightforward to show that the linear operator $\mathcal{L}$
and boundary conditions \eqref{lsa:bc}
are self-adjoint with respect to the inner product
\begin{equation}\label{eq:inner-product}
    \langle \vec{u}, \vec{u}'\rangle = \int_{-1}^1 (v v' + w w')\,\text{d}\theta.
\end{equation}
Consequently, two different eigenmodes $\check{\vec{u}}_n$ and $\check{\vec{u}}_m$ with distinct eigenvalues $\omega_n^2 \neq \omega_m^2$
that satisfy $\mathcal{L} \check{\vec{u}}_n = \omega_n^2 \check{\vec{u}}_n$
and $\mathcal{L} \check{\vec{u}}_m = \omega_m^2 \check{\vec{u}}_m$
will be orthogonal. Thus, we can normalise the eigenmodes such that
$\langle \check{\vec{u}}_n, \check{\vec{u}}_m\rangle = \delta_{nm}$, 
where $\delta_{nm}$ is the Kronecker delta.

In general, the eigenproblem \eqref{first eigenproblem}--\eqref{lsa:bc} must be solved numerically.  We do this using a finite difference discretisation.  The eigenvalues of 
the resulting sparse system are computed using
the {\tt eigs} function in SciPy.
The eigenvalues $\omega^2$ are
positive before snap-through occurs, implying the eigenmodes oscillate in time.  Snap-through occurs
when the smallest eigenvalue crosses zero
and becomes negative, resulting in 
an exponentially growing and decaying
mode.

\subsection{Symmetries of the eigenmodes}

There are  two
types of symmetries that
manifest in the eigenmodes defined by the eigenvalue problem in \eqref{first eigenproblem}
when the equilibrium solution is \emph{symmetric} and hence satisfies
\eqref{eqn:eq_symmetry}.  In particular, by making the 
substitution $\theta\to-\theta$, we find that the linear operator
$\mathcal{L}$ defined in \eqref{L operator} transforms according to
\begin{equation}\label{L symmetry eq}
    \mathcal{L} = \begin{pmatrix}
        \mathcal{L}_{vv} & \mathcal{L}_{vw} \\
        \mathcal{L}_{wv} & \mathcal{L}_{ww}
    \end{pmatrix}
    \to
    \begin{pmatrix}
        \mathcal{L}_{vv} & -\mathcal{L}_{vw} \\
        -\mathcal{L}_{wv} & \mathcal{L}_{ww}
    \end{pmatrix}.
\end{equation}
The transformation implies the existence of \emph{symmetric} eigenmodes
$\check{\vec{u}}^{S} = (\check{v}^{S}, \check{w}^S)$ that satisfy
\begin{align}
\mathcal{L}\check{\vec{u}}^S = \omega^S \check{\vec{u}}^S,
\quad
\check{v}^{S}(-\theta, t) = \check{v}^{S}(\theta, t),
\quad
\check{w}^{S}(-\theta, t) = -\check{w}^{S}(\theta, t),
\label{eqn:lin_symmetric}
\end{align}
and \emph{antisymmetric} eigenmodes $\check{\vec{u}}^{A}= (\check{v}^{A}, \check{w}^A)$ that satisfy
\begin{align}
\mathcal{L}\check{\vec{u}}^A = \omega^A\check{\vec{u}}^A,
\quad
\check{v}^{A}(-\theta, t) = -\check{v}^{A}(\theta, t),
\quad
\check{w}^{A}(-\theta, t) = \check{w}^{A}(\theta, t).
\label{eqn:lin_asymmetric}
\end{align}
If we let $\{\check{\vec{u}}^S_m\}$ and
$\{\check{\vec{u}}^A_n\}$ denote all of the
symmetric and antisymmetric eigenmodes, then
orthogonality implies that
$\langle \check{\vec{u}}^{S}_m, \check{\vec{u}}^S_{m'} \rangle = \delta_{mm'}$,
$\langle \check{\vec{u}}^{S}_m, \check{\vec{u}}^A_{n} \rangle = 0$, and
$\langle \check{\vec{u}}^{A}_n, \check{\vec{u}}^A_{n'} \rangle = \delta_{nn'}$ for all values of 
$n$, $n'$, $m$, and $m'$.

\subsection{Bifurcations and snap-through modes}
\begin{figure}
    \centering
    \includegraphics[width =1\textwidth]{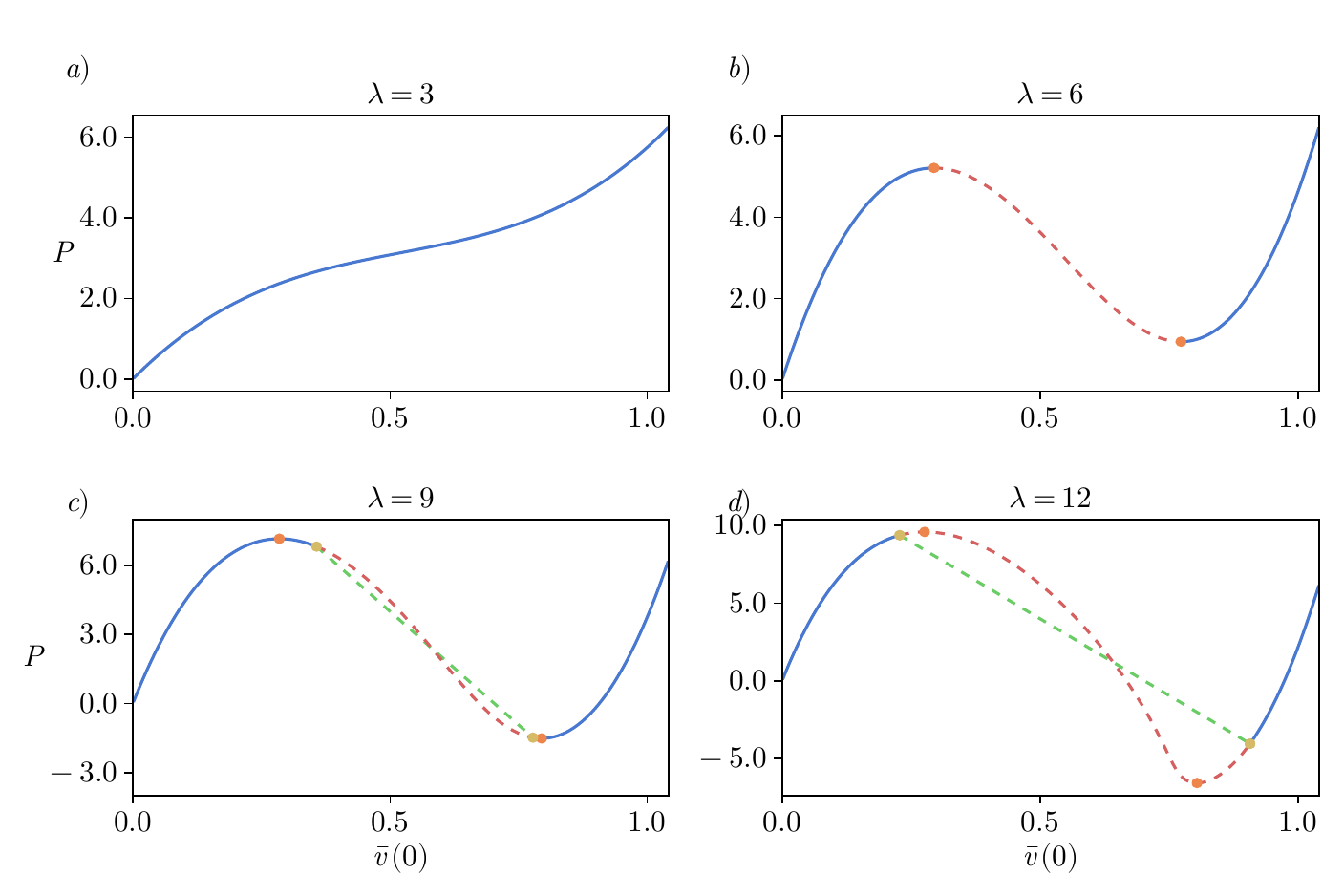}
    \caption{Vertical response of the arch midpoint $\bar{v}(0)$ to applied load $P$ for different values of $\lambda$. Blue lines denote stable symmetric equilibrium, dashed red lines denote unstable symmetric equilibrium, and dashed green lines denote a pair of unstable asymmetric solutions. Red dots indicate limit-point bifurcation and yellow dots indicate pitchfork bifurcation.}
    \label{fig:CircumferentialCompression}
\end{figure}

Snap-through occurs when a small increase
in the applied load either causes the current
stable configuration of the arch to become
unstable or cease to exist.
Mathematically, 
snap-through points correspond to bifurcations
of the equilibrium configurations and can be
found by tracking how the equilibria and their
stability evolve under the variation of the controlling parameter.
The bifurcation scenarios can be visualised by plotting how the
applied load $P$ varies with increasing
the radial displacement at the arch
mid-point $\bar{v}(0)$, as shown in Fig.~\ref{fig:CircumferentialCompression} for representative values of $\lambda$. 
We find that snap-through always emerges from a symmetric equilibrium solution, but the mode of snap-through can differ and, depending on the value of $\lambda$, it is possible to identify four cases:
\begin{enumerate}[(i)]

\item{$0 < \lambda < \lambda_c\simeq 3.9$. The applied load monotonically increases from
zero with increasing mid-point displacement (Fig.~\ref{fig:CircumferentialCompression}~(a)).  In this case, snap-through cannot
occur. The only equilibrium solution is associated with a stable symmetric configuration. }

\item{$\lambda_c < \lambda < \lambda_a \simeq 8.0$. As the midpoint displacement increases from zero, the applied load rises from zero to a maximum, then falls to a minimum, before increasing again (Fig.~\ref{fig:CircumferentialCompression}~(b)). In this case, snap-through occurs through a limit-point bifurcation of a symmetric equilibrium because, at the maximum of the curve, two symmetric equilibrium solutions of opposite stability collide and annihilate each other.  This leads to
limit-point buckling.
The value of $\lambda_c$ is given by
$\lambda_c = \sqrt{3}\pi^3/\sqrt{2\left(\pi^4-18\pi^2+48\pi+24\right)}\approx 3.905$
and it determines when the load-displacement curve has an interior inflection point with zero derivative.
The value of $\lambda_a = 6\pi^3/\sqrt{6\pi^4-9\pi^2+48}\approx7.979$ is obtained by setting $\mu=\pi$ in \eqref{static comp cond} and looking for the smallest $\lambda$ such that a real solution for $P$ exists.
}

\item{$\lambda_a < \lambda < \lambda_\text{hill}^+\simeq 10.25$. As in case (ii), the applied load increases from zero to a maximum.  However,  after the
maximum in the applied load and on the descending unstable symmetric branch, two unstable asymmetric solutions emerge via a 
pitchfork bifurcation (Fig.~\ref{fig:CircumferentialCompression}~(c)). 
Snap-through still occurs via the limit-point
bifurcation of a symmetric solution, associated with reaching the maximum of the curve (limit-point buckling). The case $\lambda=\lambda_\text{hill}^+$ corresponds to the condition for which the emergence of the two unstable asymmetric solutions coincides with the maximum of the force-displacement curve, i.e.\ it is the point at which the limit-point and pitchfork bifurcations simultaneously occur.  In 
the literature, this is known as a co-dimension 2 bifurcation called a ``hilltop-branching point" \cite{Cusp}.
The value of $\lambda_\text{hill}^{+}$ 
can be analytically computed as
\begin{gather}
    \lambda_\text{hill}^+ = \sqrt{6}\, \sqrt{\frac{34 \pi^{8}-41 \pi^{6}+96 \pi^{4}+\sqrt{29 \pi^{16}-289 \pi^{14}+1496 \pi^{12}-5520 \pi^{10}+9216 \pi^{8}}}{23 \pi^{6}-51 \pi^{4}+137 \pi^{2}-48}} \approx 10.25.
\end{gather}}

\item{$\lambda_\text{hill}^+ <\lambda $.  As the mid-point displacement increases from zero and the applied load increases, two unstable asymmetric solutions emerge via a pitchfork bifurcation before the limit-point bifurcation of symmetric solutions occurs at the maximum of the curve (Fig.~\ref{fig:CircumferentialCompression}~(d)).  At the load where the two asymmetric solutions emerge, the symmetric solution becomes unstable.  In this case, snap-through occurs via a pitchfork bifurcation, leading to bifurcation buckling.
}
\end{enumerate}
For the purpose of this analysis, we note that the equilibrium configurations at snap-through either coincide with the maximum of the force-displacement curve (limit-point bifurcation) or the point at which the asymmetric solutions are first admittable (pitchfork bifurcation), whichever happens at the smallest value of the arch midpoint displacement. We denote the configurations and loads from which snap-through occurs with a $cr$ subscript, i.e.\ $\overline{v}_{cr}$ and $P_{cr}$.

For completeness, we note that an equivalent discussion can be made by looking at the \emph{snap-back} event, i.e.\ the snapping from the remote stable configuration as the applied load is decreased. The discussion of four cases still holds by replacing the maximum of the force displacement curve with the minimum and bifurcation buckling happens for $\lambda>\lambda_{hill}^-$ given by
\begin{gather}
    \lambda_{hill}^- = \sqrt{6}\, \sqrt{\frac{34 \pi^{8}-41 \pi^{6}+96 \pi^{4}-\sqrt{29 \pi^{16}-289 \pi^{14}+1496 \pi^{12}-5520 \pi^{10}+9216 \pi^{8}}}{23 \pi^{6}-51 \pi^{4}+137 \pi^{2}-48}} \approx 9.232.
\end{gather}

\subsection{Calculation of the critical eigenmodes}
\label{sec:criticalmodes}

The snap-through configurations of the system can be
determined by seeking parameter combinations
that lead to a zero eigenvalue ($\omega = 0$)
in the eigenproblem \eqref{first eigenproblem}--\eqref{lsa:bc}.  
The critical eigenmode at snap-through associated with the zero eigenvalue is called a \emph{critical mode}, and will be denoted with a $0$ subscript.  The critical modes 
(or, equivalently, the snap-through configurations) can be obtained
by solving
\begin{equation}
    \mathcal{L}\tilde{\vec{u}}_0=\boldsymbol{0},
\end{equation}
where $\mathcal{L}$ is given by \eqref{L operator}.  However, it
is more convenient to instead calculate the critical modes by
linearising the equations that define the equilibrium solutions,
since setting $\omega = 0$ corresponds to seeking a time-independent
solution.
 We find that
$\partial \tilde{\varepsilon}_{0} / \partial \theta = 0$ and hence
the perturbation to the membrane strain, $\tilde{\varepsilon}_{0}$, is
uniform in space.  The $O(\epsilon)$ contributions to
\eqref{radial PDE} and \eqref{eq:compatibilityGeneral} reduce to
\begin{subequations}
\begin{align}\label{zero eigenvalue problem}
    &\odv[order=4]{\tilde{v}_{0}}{\theta}+\mu^2\odv[order=2]{\tilde{v}_{0}}{\theta}=\lambda^2\tilde{\varepsilon}_{0}\left(\odv[order=2]{\bar{v}}{\theta}+1\right), \\
    & \tilde{\varepsilon}_{0}=\frac{1}{2}\int^1_{-1}\left(\odv{\bar{v}}{\theta}\odv{\tilde{v}_{0}}{\theta}-\tilde{v}_{0}\right)\text{d}\theta,
    \label{zero eigenvalue comp cond}
\end{align}
\end{subequations}
respectively. Due to the possibility of
asymmetric equilibrium when $\mu = \pi$, and symmetric equilibria for any $\mu$, we must consider these two cases separately. As highlighted in the previous section, $\bar{v}$ is always symmetric and given by \eqref{v static sol}.

\subsubsection{Limit-point buckling}
If $\mu \neq \pi$, the perturbations to the displacement field are given by
\begin{subequations}\label{zero eigmodes}
\begin{align} 
    &\tilde{v}_{0}(\theta) = V^s_1+V^s_2|\theta|+(V^s_3+V^s_4|\theta|)\cos(\mu\theta)+(V^s_5+V^s_6|\theta|)\sin(\mu|\theta|), \\
    &\tilde{w}_{0}(\theta) = \iint \left(\odv{v_0}{\theta} - \odv[order=2]{\bar{v}}{\theta}\odv{v_0}{\theta}-\odv{\bar{v}}{\theta}\odv[order=2]{v_0}{\theta}\right)\text{d}\theta\,\text{d}\theta + W^s_1+W^s_2\theta,
\end{align}
\end{subequations}
with 
\begin{subequations}
\begin{align}
    V^s_1 &= \frac{2-P}{2\mu^4},
    \\
    V^s_2 &= \frac{P}{2\mu^4}, 
    \\
    V^s_3 &= \frac{3P\sin(\mu)\cos(\mu)+2\sin(\mu)\mu^2-4\mu\cos(\mu)-P\mu}{4\mu^5\cos(\mu)^2},
    \\ 
    V^s_4 &= \frac{P}{4\mu^4},
    \\
    V^s_5 &= -\frac{3P}{4\mu^5}, 
    \\
    V^s_6 &= \frac{P\sin(\mu)-2\mu}{4\mu^4\cos(\mu)},
    \\
    W^s_1 &= \frac{P\left(P\mu+8\mu\cos(\mu)-2\sin(\mu)\mu^2-5P\sin(\mu)\cos(\mu)\right)}{8\mu^7\cos(\mu)},
    \\
W^s_2 &= \frac{W^s_{21} + W^2_{22}}{32 \mu^7 \cos (\mu)^3}, \\
W^s_{21} &= 4 \mu\left(P^2-8 P+4(2-P) \mu^2\right) \cos (\mu)^3 \nonumber \\ &\quad +\left(4 P\left(\mu^2+8\right) \mu-3\left(5 P^2+2 \mu^2\right) \sin (\mu)\right) \cos (\mu)^2, \\
W^s_{22} &=  \left(P^2 \mu+4 \mu^3-4 P \mu^2 \sin (\mu)\right) \cos (\mu) +2\left(P^2 \mu^2+4 \mu^4\right) \sin (\mu)-8 P \mu^3.
\end{align}
\end{subequations}
It can be easily checked that 
\begin{align}
\tilde{v}_{0}(-\theta,t) = \tilde{v}_{0}(\theta, t), 
\qquad 
\tilde{w}_{0}(-\theta, t) = -\tilde{w}_{0}(\theta, t),
\end{align}
thus describing a symmetric critical mode. 

\subsubsection{Bifurcation buckling}
For the special case when $\mu=\pi$, the solution of eq.~\eqref{zero eigenvalue problem} implies $\tilde{\varepsilon}_{0} = 0$. The corresponding displacements are given by
\begin{subequations}\label{Other zero eigenmodes}
\begin{align}
    \tilde{v}_{0}(\theta) &= \sin(\pi\theta), \\
    \tilde{w}_{0}(\theta) &= W^a_1+W^a_2|\theta| + (W^a_3+|\theta|)\sin(\pi|\theta|)+(W_4^a\cos(\pi\theta)+W_5^a\sin(\pi|\theta|))\cos(\pi\theta),
\end{align}
\end{subequations}
where
\begin{gather}
    W_1^a = \frac{2-P}{4\pi},\quad W_2^a = \frac{P}{4\pi},\quad W_3^a = -\frac{P}{2\pi^2},\quad W_4^a = -\frac{1}{2\pi},\quad W^a_5 = \frac{P}{4\pi^2}.
\end{gather}
It can be easily checked that 
\begin{align}
\tilde{v}_{0}(-\theta,t) = -\tilde{v}_{0}(\theta, t), 
\qquad 
\tilde{w}_{0}(-\theta, t) = \tilde{w}_{0}(\theta, t),
\end{align}
thus describing an antisymmetric critical mode. 

In the following sections, we develop our theory using orthonormal eigenfunctions, so we normalize the eigenfunctions in \eqref{zero eigmodes} and \eqref{Other zero eigenmodes} as
\begin{equation}
    \check{v}_0(\theta) = \frac{\tilde{v}_0(\theta)}{\langle\tilde{\vec{u}}_0,\tilde{\vec{u}}_0\rangle},
    \qquad 
    \check{w}_0(\theta) = \frac{\tilde{w}_0(\theta)}{\langle\tilde{\vec{u}}_0,\tilde{\vec{u}}_0\rangle},
\end{equation}
where $\langle\cdot,\cdot\rangle$ is the inner product defined in \eqref{eq:inner-product}.

\section{Numerical simulations}\label{sec:Numerics}

The non-dimensional model given by
\eqref{PDEs}--\eqref{nd:ics} is numerically 
solved using the Python package FEniCS \cite{FEniCS}, which is an
open-source finite element library.  The weak
form of the governing equations is discretised
in space using 1000 nodes.  The solution is
represented using piecewise linear elements
(e.g.\ linear continuous Galerkin elements).
The system was discretised in time using the
generalised-$\alpha$ method, which is second-order accurate in time and dampens high-frequency modes~\cite{GeneralisedAlphaMethod}.
The resulting nonlinear system of algebraic
equations is solved using Newton's method
at each time step.  The sparse direct solver MUMPS was used to
solve the linear systems formed by each Newton iteration.

\section{Multiple-scales analysis of snap-through}\label{sec:MS}

To gain analytical insights into the snap-through
dynamics of the arch and to rigorously  calculate the snap-through time, we use the method of multiple scales~\cite{kevorkian1996multiple}. We assume that the system is initially at the snap-through configuration, associated with an applied load $P_{cr}$. Snap-through is then triggered by adding a small increment to the critical applied load.  Thus, we let
\begin{equation}
    P = P_{cr}(1+\DP), \quad \DP\ll 1.
\end{equation}
We then look for solutions of the form
\begin{gather}\label{eq:expansions}
    \vec{u}(\theta,t) = \vec{u}_{cr}(\theta) + \DP \hat{\vec{u}}(\theta, t),
    \qquad
    \varepsilon_m(\theta,t) = \varepsilon_{cr} + \DP\hat{\varepsilon}(\theta,t).
\end{gather}
Substituting these expressions into eq. \eqref{PDEs} and \eqref{eq:membranestrain}, we obtain the governing problem for the perturbations $\hat{\boldsymbol{u}}$ and $\hat{\varepsilon}$ as
\begin{subequations}\label{Asymptotic PDEs}
\begin{gather}
    \pdv[order=2]{\hat{v}}{t}+\frac{1}{\lambda^2}\pdv[order=4]{\hat{v}}{\theta}- \varepsilon_{cr}\pdv[order=2]{\hat{v}}{\theta} - \odv{\overline{v}_{cr}}{\theta}\pdv{\hat{\varepsilon}}{\theta}-{\hat{\varepsilon}}\left(\odv[order=2]{\overline{v}_{cr}}{\theta}+1\right)-\DP\left(\pdv{\hat{\varepsilon}}{\theta}\pdv{\hat{v}}{\theta}+{\hat{\varepsilon}} \pdv[order=2]{\hat{v}}{\theta}\right)=\frac{P_{cr}\delta(\theta)}{\lambda^2}, \\
    \pdv[order=2]{\hat{w}}{t}=\pdv{\hat{\varepsilon}}{\theta},\\{\hat{\varepsilon}} = \pdv{\hat{w}}{\theta}-{\hat{v}}+\odv{\overline{v}_{cr}}{\theta}\pdv{\hat{v}}{\theta}+\frac{1}{2}\DP\left(\pdv{\hat{v}}{\theta}\right)^2.
\end{gather}
\end{subequations}
By eliminating $\hat{\varepsilon}$, the system \eqref{Asymptotic PDEs}
can be written in a compact vector form as
\begin{equation}\label{Order 1 nonlinear eq}
     \pdv[order=2]{\hat{\boldsymbol{u}}}{t}+\mathcal{L} {\hat{\boldsymbol{u}}} = \DP \mathcal{N}({\hat{\boldsymbol{u}}})+\delta(\theta) {\boldsymbol{F}},
\end{equation}
where $\mathcal{L}$ is the linear differential operator defined by \eqref{L operator}, $\mathcal{N}$ is a nonlinear differential operator, and $\vec{F} = (P_{cr} / \lambda^2, 0)$.  The initial conditions in eq.~\eqref{nd:ics} become
\begin{equation}\label{eq:ic_hat}
    \hat{\vec{u}}(\theta,0) = 0, \quad \frac{\partial  \hat{\vec{u}}}{\partial t}(\theta,0)  = 0.
\end{equation}
The
operator $\mathcal{N}$ can be expressed as
the sum of a bilinear and trilinear operator,
\begin{align}
\mathcal{N}(\vec{u}) = \mathcal{B}(\vec{u}, \vec{u}) + \DP\, \mathcal{T}(\vec{u}, \vec{u}, \vec{u}),
\end{align}
where 
\begin{align}
\mathcal{B}(\vec{u}, \vec{u}) = \begin{pmatrix}
        3\odv{\overline{v}_{cr}}{\theta}\pdv{v}{\theta}\pdv[order=2]{v}{\theta}+\frac{1}{2}\left(3\odv[order=2]{v _{cr}}{\theta}-1\right)\left(\pdv{v}{\theta}\right)^2-v\pdv{v}{\theta}+\pdv{v}{\theta}\pdv[order=2]{w}{\theta}+\pdv[order=2]{v}{\theta}\pdv{w}{\theta} \\ \pdv{v}{\theta}\pdv[order=2]{v}{\theta}
    \end{pmatrix}
\end{align}
and
\begin{align}
\mathcal{T}(\vec{u}, \vec{u}, \vec{u})  = \begin{pmatrix}
        \frac{3}{2}\left(\pdv{v}{\theta}\right)^2\pdv[order=2]{v}{\theta} \\ 0
    \end{pmatrix}.
\end{align}

By using the symmetries of the modes of $\mathcal{L}$ discussed in Sec.~\ref{sec:lsa}, we write the solution to \eqref{Order 1 nonlinear eq} in terms of an eigenfunction expansion decomposed into symmetric and antisymmetric eigenmodes as
\begin{align}
\hat{\vec{u}}(\theta, t) = 
\sum_{n \in \mathbb{S}} S_n(t) \check{\vec{u}}_n^S(\theta) + 
\sum_{m \in \mathbb{A}} A_m(t) \check{\vec{u}}_m^A(\theta),
\label{eqn:u_expansion}
\end{align}
where superscript $S$ indicates symmetric eigenmodes, superscript $A$ indicates antisymmetric eigenmodes,  $S_n$ and $A_m$ are amplitudes of symmetric and antisymmetric modes, respectively, and $\mathbb{S}$ and $\mathbb{A}$ are sets that contain
the indices of the symmetric and antisymmetric modes (which will be 
specified when pertinent).

Substituting \eqref{eqn:u_expansion} into \eqref{Order 1 nonlinear eq} and then
taking inner products with symmetric and antisymmetric eigenmodes
leads to a nonlinear system of equations for the amplitudes
given by
\begin{subequations}
\label{eqn:ode_as}
\begin{align}
\tdd{S_n}{t} + (\omega_n^S)^2 S_n &= F_n +\DP\left\langle\mathcal{N}(\hat{\vec{u}}),\check{\vec{u}}_n^S\right\rangle,  \\
\tdd{A_m}{t} + (\omega_m^A)^2 A_m &=  \DP\left\langle\mathcal{N}(\hat{\vec{u}}),\check{\vec{u}}_m^A\right\rangle,
\end{align}
\end{subequations}
where $F_n = \langle \vec{F}\delta(\theta), \check{\vec{u}}^S_n \rangle$. From \eqref{eqn:ode_as}, we see that the symmetric nature of the point load only directly excites the symmetric modes. 
Growth of antisymmetric modes can only happen 
via non-linear interactions.
When written using the same eigenfunction expansion, the initial conditions in eq.~\eqref{eq:ic_hat} become
\begin{align}\label{eq:ic_as}
S_n(0) = 0, \quad \td{S_n}{t}(0) = 0, \quad A_m(0) = 0, \quad \td{A_m}{t}(0) = 0.
\end{align}
These conditions represent an ideal scenario where
snap-through happens from exactly the equilibrium configuration that becomes unstable at the bifurcation.  In our analysis, we highlight how ideal initial conditions lead to very different behaviour when applied to the limit-point and bifurcation buckling.

For convenience, we introduce the quantities obtained by taking the inner product between the bilinear and trilinear forms and the eigenmodes of $\mathcal{L}$:
\begin{equation}
    \N{nmk}^{IJK} := \langle \mathcal{B}(\check{\vec{u}}^I_n,\check{\vec{u}}^J_m),\check{\vec{u}}^K_k \rangle,
    \qquad 
    \Nt{nmsk}^{IJKL} := \langle \mathcal{T}(\check{\vec{u}}^I_n,\check{\vec{u}}^J_m,\check{\vec{u}}^K_s),\check{\vec{u}}^L_k \rangle,
\end{equation} 
where the superscripts $I,J,K,L\in\{A,S\}$ are used to identify the symmetries of the eigenmodes denoted by the subscripts $n,m,k,s$. Due of the symmetries in the problem, it can be shown that
\begin{subequations}
\label{eqn:sym_simp}
\begin{align}
\mathcal{B}_{nmk}^{SSA} = \left\langle\mathcal{B}\left(\check{\vec{u}}_n^S,\check{\vec{u}}_{m}^S\right),\check{\vec{u}}_k^A\right\rangle=0,\\
\mathcal{T}_{nmks}^{SSSA} = \left\langle\mathcal{T}\left(\check{\vec{u}}_n^S,\check{\vec{u}}_{m}^S,\check{\vec{u}}_{k}^S\right),\check{\vec{u}}_s^A\right\rangle=0,
\end{align}
\end{subequations}
which will be used in the analysis.

\subsection{Analysis of limit-point buckling}\label{sec:LP}

Limit-point buckling occurs when the critical eigenvalue $\omega_0=0$ is associated with a symmetric eigenmode $\check{\vec{u}}_0=\check{\vec{u}}_0^S$. Without loss of generality, we rearrange the remaining eigenvalues in increasing order as $0 < \omega_1^S < \omega_2^S < \ldots$ and $0 < \omega_1^A < \omega_2^A < \ldots$, 
implying that $\mathbb{S} = \{0, 1, 2, \ldots\}$
and $\mathbb{A} = \{1, 2, \ldots \}$.

\subsubsection{The origin of multiple time scales}
We first consider a standard asymptotic expansion of the amplitudes of the symmetric and antisymmetric modes as
\begin{subequations}
\label{eq:wnl_expansion}
\begin{align}
    S_n(t)&=\sum_{k=0}^{\infty}\DP^k S_n^{(k)}(t), \\
    A_m(t)&=\sum_{k=0}^{\infty}\DP^k A_m^{(k)}(t).
\end{align}
\end{subequations}
After inserting \eqref{eq:wnl_expansion} in eq.~\eqref{eqn:ode_as}, with initial conditions in eq.~\eqref{eq:ic_as}, we then expand in powers of $\Delta P$.  By collecting terms at each order of $\Delta P$, a hierarchy of linear systems of ordinary differential equations can be formulated and then solved.

From the $O(1)$ problem, we obtain the solution
\begin{subequations} \label{Sym sol}
\begin{alignat}{2}\label{Sym quad sol}
    \St{0}{0}{t} &= F_0t^2,\\
    \St{n}{0}{t} &= \frac{F_n}{(\omega^S_n)^2}\left(1-\cos(\omega_n^St)\right),\\
    \A{n}{0}{t} &= 0,
\end{alignat}
\end{subequations}
for $n = 1, 2, \ldots$. 
In deriving \eqref{Sym sol}, it was assumed that the 
nonlinear term $\DP\left(\St{0}{0}\right)^2$ remains
small.  Due to the quadratic growth of $\St{0}{0}$,
this nonlinear term will become $O(1)$ in size when $t = O(\DP^{-1/4})$, at which point the amplitude of the critical mode is given by $\St{0}{0} = O(\DP^{-1/2})$ and the solution \eqref{Sym quad sol} is no longer valid.
By resolving the solution on $O(1)$ time scales using asymptotic methods, we have systematically derived the scales needed to capture the nonlinear (snap-through) behaviour, which are the same as those proposed ad hoc for an equivalent snap-through problem in \cite{Gomez2016} and confirmed numerically in \cite{Radisson_2023}.

\subsubsection{Multiple-scale expansion}
\begin{figure}
    \centering
    \includegraphics[width=0.8\linewidth]{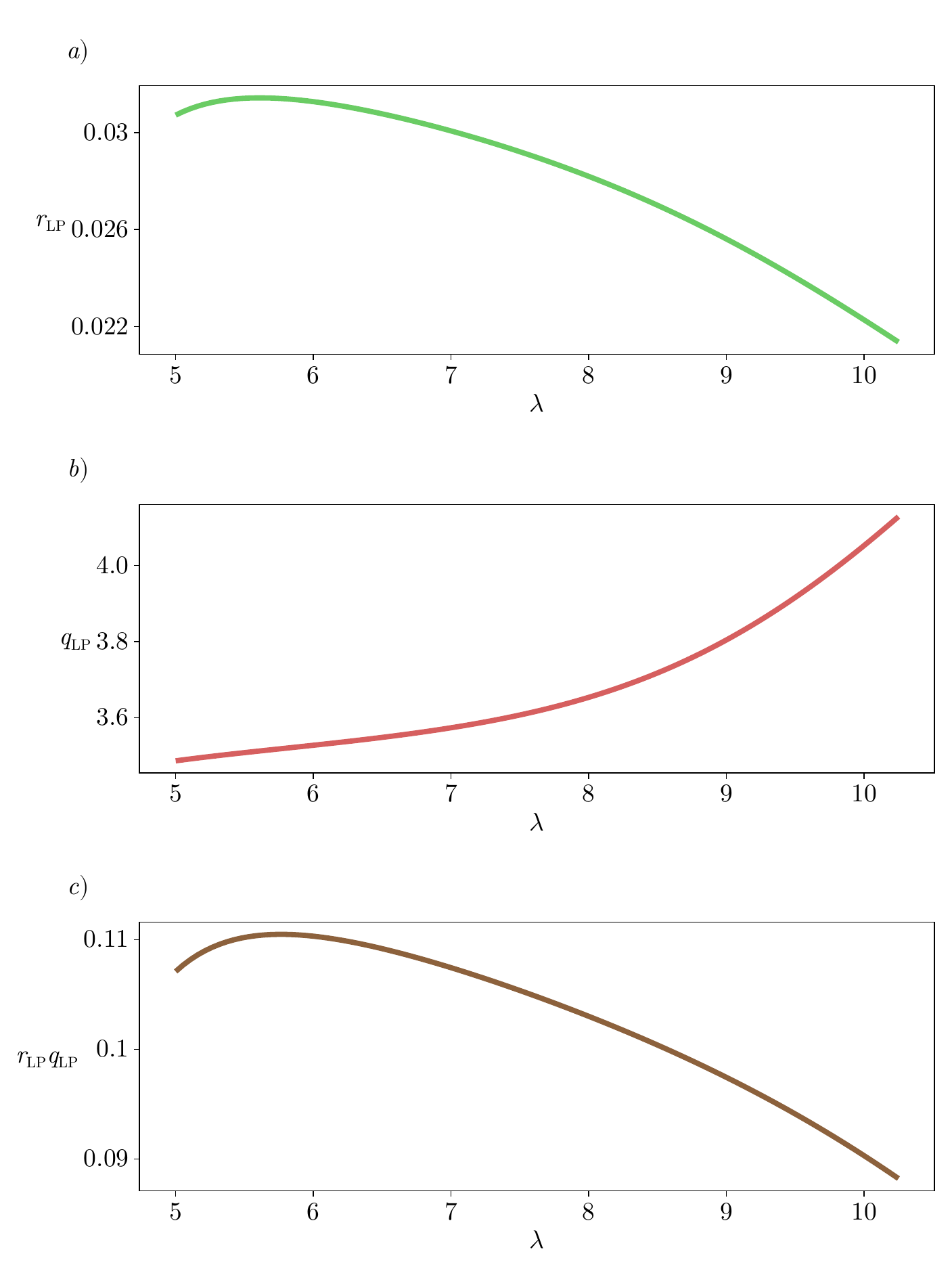}
    \caption{Quantities (a) $r_\text{LP}$, (b) $q_\text{LP}$ and (c) the product $r_\text{LP}q_\text{LP}$ for the interval of $\lambda$ where the arch undergoes limit-point buckling.}
    \label{fig:qr_limitpoint}
\end{figure}

To simultaneously account for the oscillatory dynamics on an $O(1)$ time scale and the nonlinear growth of the
critical mode on an $O(\DP^{-1/4})$ time scale, 
we introduce a new slow time variable given by
$\tau = \DP^{1/4}t$.
Moreover, we expand the amplitudes as
$S_0 \sim \DP^{-1/2}\bar{S}_0^{(0)}(\tau) + \bar{S}_{0}^{(1)}(t, \tau) + \ldots$, 
$S_n \sim \bar{S}_n^{(0)}(t, \tau) + \ldots$ and
$A_n \sim \bar{A}_n^{(0)}(t, \tau) + \ldots$ for
$n = 1, 2, \ldots$.
The leading-order contribution to the evolution 
equation for the critical amplitude is
\begin{equation}
     \pdv[order=2]{\bar{S}_0^{(1)}}{t} =  \left(\bar{S}_0^{(0)}\right)^2 \mathcal{B}^{SSS}_{000}+F_0 - \odv[order=2]{\bar{S}_0^{(0)}}{\tau},
\end{equation}
with homogeneous initial conditions. To suppress secular terms, we set the right-hand side (RHS) to zero to obtain 
\begin{equation}\label{Sym amplitude eq}
    \odv[order=2]{\bar{S}_0^{(0)}}{\tau} = r_{\text{LP}}+q_{\text{LP}} \left(\bar{S}_0^{(0)}\right)^2,
\end{equation}
where
\begin{equation}
    r_{\text{LP}} = F_0,\wand q_{\text{LP}} =\N{000}^{SSS}.
\end{equation}
The implict solution to \eqref{Sym amplitude eq} with homogeneous initial conditions is given by
\begin{equation}
    \tau = \sqrt{\frac{3}{2}}\int_0^{\bar{S}_0^{(0)}(\tau)} \frac{\text{d}\xi}{\sqrt{3r_{\text{LP}}\xi+q_{\text{LP}}\xi^3}}.
\end{equation}
By taking the limit for $\bar{S}_0^{(0)}\to\infty$, the finite snap-through time can be computed as
\begin{equation}
    \tau_\text{snap} 
    = \left(\frac{3}{64\pi^2r_{\text{LP}}q_{\text{LP}}}\right)^\frac{1}{4}\Gamma\left(\frac{1}{4}\right)^2,
\end{equation}
or, in the original (non-dimensional) time variable, as
\begin{equation}\label{Symmetric snap-through time}
    t_\text{snap}^{\text{LP}} = \left(\frac{3}{64\pi^2r_{\text{LP}}q_{\text{LP}}\DP}\right)^\frac{1}{4}\Gamma\left(\frac{1}{4}\right)^2.
\end{equation}
The limit-point snap-through time given by \eqref{Symmetric snap-through time} is equivalent to the one derived in \cite{Gomez2016} for the case of the compressed beam. Because the critical mode $\check{\vec{u}}^S_0 $ can be found analytically, the coefficients $r_{\text{LP}}$ and $q_{\text{LP}}$ can be derived in an exact form and their dependence on $\lambda$ can be found explicitly. As shown in Fig. \ref{fig:qr_limitpoint}, the quantity $r_{\text{LP}}q_{\text{LP}}$ remains approximately constant in the region where the arch undergoes limit-point buckling. The expression in \eqref{Symmetric snap-through time} is thus independent of $\lambda$, as discussed in the main text. 
In this mode of snap-through, the load acts directly on the critical mode via $F_0$, which ensures that snap-through occurs even when considering the ideal initial conditions.

\subsection{Analysis of bifurcation buckling}\label{sec:BifBuck}
Bifurcation buckling occurs when the critical eigenvalue $\omega_0=0$ is associated with a antisymmetric eigenmode $\check{\vec{u}}_0=\check{\vec{u}}_0^A$. We rearrange the remaining eigenvalues as $0 < \omega_1^S < \omega_2^S < \ldots$ and $0 < \omega_1^A < \omega_2^A < \ldots$.  The sets of indices are defined as
$\mathbb{A} = \{0, 1, 2, \ldots\}$
and $\mathbb{S} = \{1, 2, \ldots \}$. In this case, we first show that imperfections in the initial conditions are required to trigger snap-through. We then discuss the structure of the multiple-scale analysis in the presence of the revised initial conditions.\par

\subsubsection{Elucidating the need for imperfections}\label{sec:Imperfections}
We consider again an asymptotic expansion of the amplitudes of the symmetric and antisymmetric modes given by \eqref{eq:wnl_expansion}.
After inserting this ansatz in eq.~\eqref{eqn:ode_as}, with initial conditions in eq.~\eqref{eq:ic_as}, we obtain, at $O(1)$, the amplitudes
\multia[A01 solution unperturbed]{
    \A{0}{0}{t} &=\A{n}{0}{t}= 0,\\
    \St{n}{0}{t} &=\frac{F_n}{(\omega_n^S)^2}\left(1-\cos(\omega_n^St)\right), \label{A01 solution perturbed oscillation}
}
for $n = 1, 2, \ldots$.  Thus, we see that the
critical amplitude $\A{0}{0}$ does not grow in time
at this order.  Only the symmetric modes are
non-zero because they are excited by the
(symmetric) point load. 

Proceeding to the next order, we find that 
the evolution equation for the correction
to the critical amplitude,  $\A{0}{1}$, is given by
\begin{equation}
    \odv[order=2]{\A{0}{1}}{t} =\left\langle \mathcal{B}\left(\sum_{n\in\mathbb{S}}\St{n}{0}{t}\check{\vec{u}}_n^S,\sum_{n'\in\mathbb{S}}\St{n'}{0}{t}\check{\vec{u}}_{n}'^S\right),\check{\vec{u}}^A_0 \right\rangle= \sum_{n,n'\in\mathbb{S}}\St{n}{0}{t}\St{n'}{0}{t}\mathcal{B}^{SSA}_{nn'0}=0.\label{unperturbed A eq order 2}
\end{equation}
Solving this equation with homogeneous initial conditions shows that $\A{0}{1}=0$ necessarily. 
Similarly, it can be shown that all the antisymmetric 
amplitudes are zero at this order,
$\A{n}{1} = 0$.  However, the amplitudes of
the symmetric modes, $\St{n}{1}$, are non-zero
due to nonlinear forcing from the leading-order solution.

By proceeding to higher orders and using orthogonality of eigenfunctions, we can show that the critical 
amplitude does not grow at any order, 
$\A{0}{n}=0$ for all $n\geq0$.  Analogously, none of the amplitudes of the antisymmetric modes
grow. Thus,
nonlinear interactions between the modes are
not sufficient to trigger snap-through when
ideal initial conditions are used and we expect that the system oscillates indefinitely in a symmetric configuration.  

The analysis of the problem under ideal initial conditions reveals the need to introduce a symmetry-breaking mechanism that excites the antisymmetric modes.  While there are various ways of doing this, e.g.\ accounting for heterogeneities in the material properties or a non-centred point load, we choose to introduce an imperfection in the initial conditions.  We thus replace the ideal initial conditions \eqref{eq:ic_hat} with 
\begin{equation}\label{eq:icimpefection_hat}
    \hat{\vec{u}}(\theta,0) = \eta \check{\vec{u}}_{p}(\theta), \quad \frac{\partial  \hat{\vec{u}}}{\partial t}(\theta,0)  = \vec{0}
\end{equation}
where $\check{\vec{u}}_{p}(\theta)$ and $\eta$ capture the shape and magnitude of the imperfection. To streamline the presentation, we will assume that $\eta = O(\DP^\alpha)$, where $\alpha \geq 1$ is an integer.  However, an equivalent analysis can be carried out for an arbitrarily sized $\eta$. The requirement that $\alpha \geq 1$ ensures that the imperfection being applied to the system is small and does not affect the linear operator $\mathcal{L}$.

The solution is again written as the eigenfunction expansion posed in eq.~\eqref{eqn:u_expansion} and we asymptotically expand the amplitudes as in \eqref{eq:wnl_expansion}.
The initial conditions at each order of $\DP$ are
as follows.  For $k \neq \alpha$, we have
\begin{align}
\St{n}{k}{0} = 0,
\quad
\A{m}{k}{0} = 0, 
\quad
\odv{\St{n}{k}}{t}(0) = 0, 
\quad
\odv{\A{m}{k}}{t}(0) = 0,
\end{align}
for $n \in \mathbb{S}$ and $m \in \mathbb{A}$.
For $k = \alpha$, we have
\begin{align}
\St{n}{\alpha}{0} = \langle\check{\vec{u}}_{p},\check{\vec{u}}_n^S \rangle \equiv \check{p}^S_n,
\quad
\A{m}{\alpha}{0} = \langle\check{\vec{u}}_{p},\check{\vec{u}}_m^A \rangle \equiv \check{p}^A_m, 
\quad
\odv{\St{n}{\alpha}}{t}(0) = 0, 
\quad
\odv{\A{m}{\alpha}}{t}(0) = 0,
\label{IC from perturbation}
\end{align}
for $n \in \mathbb{S}$ and $m \in \mathbb{A}$.

The problems from $O(1)$ up to and including $O(\DP^{\alpha-1})$ are the
same as those already discussed in the case of ideal initial conditions.  Thus, we find that
\begin{subequations}
\label{eqn:as_as0}
\begin{align}
\A{m}{0}(t) &= 0, \\
\St{n}{0}{t} &=\frac{F_n}{(\omega_n^S)^2}\left(1-\cos(\omega_n^St)\right),
\label{A01 solution perturbed oscillation_v2}
\end{align}
\end{subequations}
for all $m \in \mathbb{A}$ and $n \in \mathbb{S}$.
Furthermore, we obtain
$\A{m}{k}(t) = 0$ and $\St{n}{k}(t) \neq 0$, where
$k = 1, \ldots, \alpha - 1$.  The specific form
of $\St{n}{k}$ is not needed for the rest of the analysis but we highlight that these amplitudes are non-zero.
At $O(\DP^\alpha)$, the antisymmetric amplitudes become non-zero and are given by
\begin{subequations}
\label{asy_amp}
\begin{align}
\A{0}{\alpha}{t} &= \check{p}^A_0, \\
\A{m}{\alpha}{t} &= \check{p}^A_m\cos(\omega_m^At),
\end{align}
\end{subequations}
for $m = 1, 2, \ldots$. 

To proceed to the $O(\DP^{\alpha+1})$ problem,
we recognise
that the RHS of the evolution equation for the amplitude $\A{0}{\alpha+1}$ contains resonant terms due to either: (i) the interactions of all the symmetric terms at the previous orders (with the correct size via the bilinear and trilinear forms) or (ii) the interaction between the leading-order symmetric term \eqref{A01 solution perturbed oscillation_v2} and the leading-order antisymmetric terms \eqref{asy_amp} via the bilinear form $\mathcal{B}$. 
The former set of interactions does not contribute
to the growth of $\A{0}{\alpha+1}$ because they
enter via the operators in \eqref{eqn:sym_simp}
which must be zero. The latter interactions, however, do contribute to the growth of
$\A{0}{\alpha+1}$ and lead to the following
evolution equation
\begin{align}\label{First A a+1 eq}
\dA{2}{0}{\alpha+1}{t} = 
2\check{p}^A_0\sum_{n\in\mathbb{S}}\frac{F_n \mathcal{B}^{SAA}_{n00}}{(\omega_n^S)^2}\left(1-\cos(\omega_n^St)\right)+2\sum_{\substack{n\in\mathbb{S}\\m\in\mathbb{A}}}\frac{F_n\check{p}^A_m\mathcal{B}^{SAA}_{nm0}}{(\omega_n^S)^2}\left(1-\cos(\omega_n^St)\right)\cos(\omega_m^At),
\end{align}
which is to be solved with homogeneous initial conditions.

The solution to \eqref{First A a+1 eq} grows
quadratically in time, lead to the large-time behaviour $\A{0}{\alpha+1} = O(t^2)$ as $t \to \infty$.
The asymptotic solution 
breaks down when the nonlinear interaction between $\St{n}{0}$ and $\A{0}{\alpha+1}$, appearing at the following order, is no longer small, i.e.\ when $\DP \St{n}{0} \A{0}{\alpha+1} = O(1)$, implying $t = O(\DP^{-1/2})$ since $\St{n}{0} = O(1)$ for all
time. The emergence of this time scale, proposed numerically in \cite{Radisson_2023}, is here formally justified by the necessary presence of an initial imperfection that contains antisymmetric contributions that can trigger the growth of the antisymmetric critical mode. 
Importantly, the size of the imperfection, which
is controlled by the parameter $\alpha$, does not
play a role in selecting this time scale,
thus making it robust with respect to the size of the imperfection.

\subsubsection{Multiple-scales expansion}\label{sec:BifBuck2}

To simultaneously account for oscillatory behaviour
of the symmetric modes on an $O(1)$ time scale along
with the growth of the critical antisymmetric 
mode on an $O(\DP^{-1/2})$ time scale, we 
seek a solution in the form of a multiple-scales
expansion.  We thus introduce an additional
time variable as $\tau = \DP^{1/2} t$. 
The amplitudes of the modes are expanded as
\begin{subequations}\label{MS expansion antisymmetric}
\begin{align}
A_0(t,\tau) &\sim \DP^{\alpha}[\bar{A}_0^{(\alpha)}(\tau) + \DP^{1/2}\bar{A}_{0}^{(\alpha+1/2)}(\tau)
+ \DP \bar{A}_0^{(\alpha+1)}(t, \tau) + \ldots],
\\
A_n(t,\tau) &\sim \DP^{\alpha}[\bar{A}_n^{(\alpha)}(t,\tau) + \DP^{1/2} \bar{A}_n^{(\alpha+1/2)}(t,\tau) + \ldots], 
\\ 
S_{n}(t,\tau) &\sim \bar{S}_n^{(0)}(t,\tau) + 
\DP^{1/2} \bar{S}_n^{(1/2)}(t,\tau) + \ldots
\end{align}
\end{subequations}
for $n = 1, 2, \ldots$.  
By considering the $O(1)$ and $O(\DP^{1/2})$ problems
for the symmetric modes, we can show that the
solution for the symmetric amplitudes $\bar{S}^{(0)}_n$ is given by
\eqref{A01 solution perturbed oscillation_v2}.  Similarly, the solution for the 
non-critical antisymmetric amplitudes is 
\begin{align}
    \bar{A}_m^{(\alpha)}(t,\tau) = \check{p}^A_m\cos(\omega_m^At).
\end{align}
To determine $\bar{A}_{0}^{(\alpha)}$, we consider the $O(\DP^{\alpha+1})$
problem for the critical antisymmetric amplitude.  By employing \eqref{eqn:sym_simp} and \eqref{A01 solution perturbed oscillation_v2},
we obtain
\multia[Second A a+1 eq]{
   \pdv[order=2]{\bar{A}^{(\alpha+1)}_0}{t} = 
   -\odv[order=2]{\bar{A}_{0}^{(\alpha)}}{\tau} + \left(2\sum_{n\in\mathbb{S}}\frac{F_n\N{n00}}{(\omega_n^S)^2}\left(1-\cos(\omega_n^St)\right)\right)\bar{A}_{0}^{(\alpha)}.
}
Eliminating secular terms leads to the
following initial value problem
\begin{equation}
 \odv[order=2]{\bar{A}_{0}^{(\alpha)}}{\tau} -\check{r}_\text{B}\bar{A}_{0}^{(\alpha)}=0,\quad \bar{A}_{0}^{(\alpha)}(0) = \check{p}^A_0,\quad \odv{\bar{A}_{0}^{(\alpha)}}{\tau}(0)=0,
 \label{lin_A}
\end{equation}
where 
\begin{equation}\label{eq:ra}
    \check{r}_\text{B}= 2\sum_{n\in\mathbb{S}}\frac{F_n}{(\omega_n^S)^2}\N{n00}^{SAA}.
\end{equation}
Therefore, the (slow) growth of the amplitude of the critical mode is described by
\begin{equation}
    \bar{A}_{0}^{(\alpha)} = \frac{\check{p}^A_0}{2}\left(\e^{\sqrt{\check{r}_\text{B}}\tau}+\e^{-\sqrt{\check{r}_\text{B}}\tau}\right).\label{First A 0 sol}
\end{equation}
Unlike the case of limit-point buckling, where eq.~\eqref{Sym amplitude eq} shows that the
critical mode initially grows quadratically in time
(making it known as ``algebraic snap-through"), 
eq.~\eqref{First A 0 sol} shows that
bifurcation buckling leads to exponential
growth of the critical mode.  Consequently,
bifurcation buckling
is sometimes referred to as ``exponential snap-through" \cite{Radisson_2023}.
The term $\check{r}_\text{B}$ in \eqref{First A 0 sol} represents the rate of the exponential growth.  By examining
eq.~\eqref{eq:ra}, we see that this exponential growth originates from the
the nonlinear interaction between the leading-order symmetric modes, which oscillate on $O(1)$ time
scales with
amplitude given by eq.~\eqref{A01 solution perturbed oscillation_v2}, and the initial imperfection. 
Moreover, it is this interaction which initiates
snap-through on $O(\DP^{-1/2})$ time scales.

An examination of \eqref{lin_A} shows that, 
on $O(\DP^{-1/2})$ time scales, the critical 
amplitude evolves due to linear dynamics.  The
resulting exponential growth will eventually
promote nonlinear terms that impact the subsequent
evolution of the system.  By examining the
terms that have been neglected, we find that
the multiple-scales solution breaks down when
the quadratic interactions between the critical
antisymmetric modes enters the leading-order problem
for the symmetric modes, which occurs when
$\DP\mathcal{B}(\DP^\alpha\bar{A}_{0}^{(\alpha)},\DP^\alpha\bar{A}_{0}^{(\alpha)}) = O(1)$ or, equivalently, $\bar{A}_0^{(\alpha)} = O(\DP^{-\alpha - 1/2})$.
Given that 
\begin{align}
\bar{A}_{0}^{(\alpha)} \sim \frac{\check{p}^A_0}{2}\,\e^{\sqrt{\check{r}_\text{B}}\tau}
\label{exp_match}
\end{align}
as $\tau \to \infty$, it follows that the breakdown
of the asymptotic solution occurs when
\begin{equation}\label{ssss}
\tau = O\left(\ln(\DP^{-\alpha - 1/2})\right) = O\left(\ln(1/\DP)\right)
\end{equation}
and ${A}_{0} = O(\DP^{-1/2})$. 
This results motivates us to introduce a new
slow time variable
\begin{equation}\label{tau eq}
    \tilde{\tau} = \DP^{1/2}t-\frac{1}{\sqrt{\check{r}_\text{B}}}\ln\left( \DP^{-\alpha-1/2}\right)
\end{equation}
and to seek a multiple-scales expansion of
the form $A_0 \sim \DP^{-1/2}[\tilde{A}_0^{(0)}(\tilde{\tau}) + \DP^{1/2}\tilde{A}_0^{(1/2)}(\tilde{\tau}) + \DP\tilde{A}_{0}^{(1)}(t,\tilde{\tau}) + \ldots]$ and $S_n \sim \tilde{S}^{(0)}_n(t,\tilde{\tau}) + \DP^{1/2} \tilde{S}_n^{(1/2)}(t, \tilde{\tau}) + \ldots$ for
$n \in \mathbb{S}$.  The non-critical amplitudes
$A_m$ for $m = 1, 2, \ldots$ do not impact
the leading-order behaviour of the critical
amplitude $A_0$ in this time regime so we do not
consider these.

\begin{figure}
    \centering
    \includegraphics[width=0.8\linewidth]{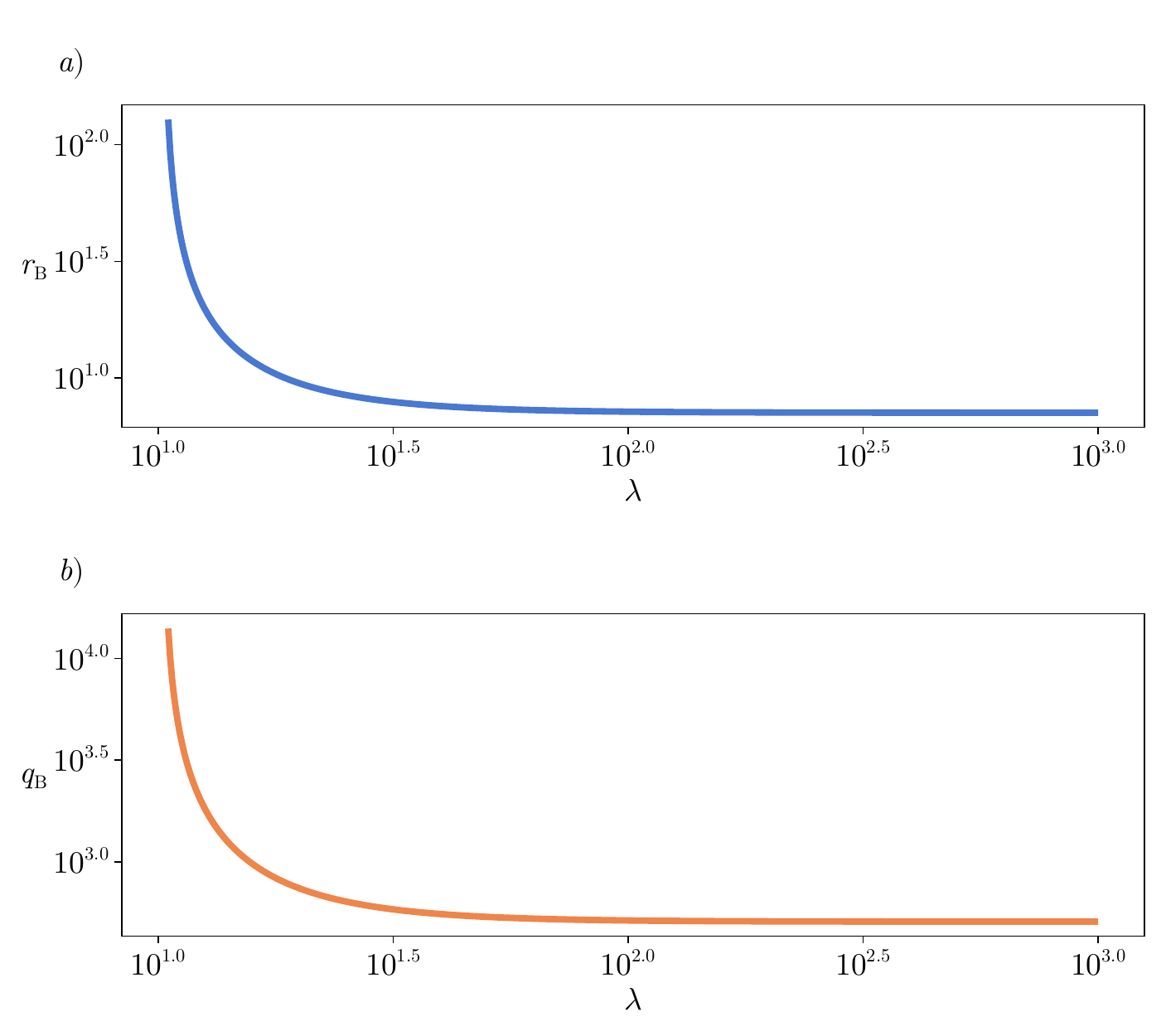}
    \caption{Quantities (a) $r_\text{B}$ and (b) $q_\text{B}$ for the interval of $\lambda$ where the arch undergoes bifurcation buckling.}
    \label{fig:rqB}
\end{figure}

From the $O(1)$ problem, we obtain
\begin{align}
     \pdv[order=2]{\tilde{S}_{n}^{(0)}}{t} + (\omega_n^S)^2 \tilde{S}_{n}^{(0)} &=  \mathcal{B}^{AAS}_{00n}[\tilde{A}_{n}^{(0)}(\tilde{\tau})]^2 +F_n,
\end{align}
which has the general solution
\begin{equation}
    \tilde{S}_n^{(0)} = \frac{F_n}{\left(\omega_n^S\right)^2}+\frac{\mathcal{B}^{AAS}_{00n}}{\left(\omega_n^S\right)^2}\,[\tilde{A}_{n}^{(0)}(\tilde{\tau})]^2+c_1(\tilde\tau) \cos\left(\omega_n^S t\right)+c_2(\tilde\tau) \sin\left(\omega_n^S t\right).
\end{equation}
The $O(\DP^{1/2})$ problem for the symmetric 
amplitudes then shows that $c_1$ and $c_2$ are
constants.  By matching to \eqref{A01 solution perturbed oscillation_v2} as $\tilde{\tau} \to -\infty$, we can deduce that
$c_1 = -F_n / (\omega_n^S)^2$ and $c_2 = 0$.  
Thus, the symmetric amplitudes are given by
 \begin{equation}
     \tilde{S}_n^{(0)} = \frac{F_n}{\left(\omega_n^S\right)^2}[1 - \cos(\omega_n^S t)] + 
     \frac{\mathcal{B}^{AAS}_{00n}}{\left(\omega_n^S\right)^2}\,[\tilde{A}_{n}^{(0)}(\tilde{\tau})]^2.
\end{equation}
At order $O(\DP^{1/2})$ we obtain
\begin{align}
     \pdv[order=2]{\tilde{A}_0^{(1)}}{t}  = -\odv[order=2]{\tilde{A}_0^{(0)}}{\tilde{\tau}} +\sum_{n\in\mathbb{S}} \frac{2}{(\omega_n^S)^2}\mathcal{B}^{SAA}_{n00}\left[F_n\tilde{A}_0^{(0)}\left(1-\cos(\omega_n^St)\right)+\mathcal{B}^{AAS}_{00n}\left(\tilde{A}_0^{(0)}\right)^3\right]+ \mathcal{T}^{AAAA}_{0000}\left(\tilde{A}_0^{(0)}\right)^3.
\end{align}
To guarantee the consistency of our expansion, we must impose that the terms independent of $t$ in the RHS are zero and thus we obtain a governing equation for the critical amplitude $\tilde{A}_{0}^{(0)}$ as
\begin{equation}\label{final amplitude eq}
    \odv[order=2]{\tilde{A}_{0}^{(0)}}{\tilde{\tau}} = \check{r}_\text{B}\tilde{A}_{0}^{(0)} + \check{q}_\text{B}\left(\tilde{A}_{0}^{(0)}\right)^3,
\end{equation}
where
\multig[asymmetric params]{
\check{r}_\text{B}= 2\sum_{n\in\mathbb{S}} \frac{\mathcal{B}^{SAA}_{n00}F_n}{(\omega_n^S)^2},\\
\check{q}_\text{B} = \mathcal{T}_{0000}^{AAAA}+2\sum_{n\in\mathbb{S}}\frac{\mathcal{B}^{SAA}_{n00}\mathcal{B}^{AAS}_{00n} }{(\omega_n^S)^2}  .
}
Eq. \eqref{final amplitude eq} can be integrated once to
obtain
\begin{align}
\odv[order=1]{\tilde{A}_{0}^{(0)}}{\tilde{\tau}} = \left[\check{r}_\text{B}\left(\tilde{A}_{0}^{(0)}\right)^2 + \frac{\check{q}_\text{B}}{2}\left(\tilde{A}_{0}^{(0)}\right)^4 + C\right]^{1/2},
\label{first_int}
\end{align}
where $C$ is a constant of integration.  Since asymptotic matching requires
$\tilde{A}_0^{(0)} \sim (\check{p}^A_0/2) \e^{\sqrt{\check{r}_\text{B}}\tilde{\tau}}$ as $\tilde{\tau} \to -\infty$ (see \eqref{exp_match}), we can substitute 
this into \eqref{first_int} and take the limit $\tilde{\tau} \to -\infty$ to find
that $C = 0$.  Integration
of \eqref{first_int} then leads to
\begin{align}
\tilde{A}_0^{(0)} = \sqrt{\frac{8 \check{r}_B}{\check{q}_B}} \frac{K \e^{\sqrt{\check{r}_B}\tilde{\tau}}}{1 - K^2 \e^{2 \sqrt{\check{r}_B} \tilde{\tau}}},
\label{eqn:A_K}
\end{align}
where $K$ is another constant of integration.  Applying
the matching condition again enables $K$ to be determined as
\begin{align}
K = \frac{\check{p}_0^A}{2} \sqrt{\frac{\check{q}_B}{8 \check{r}_B}}.
\label{K_soln}
\end{align}
The snap-through time for bifurcation buckling can then
be determined as the time at which the solution \eqref{eqn:A_K} blows up, leading to
\begin{equation}\label{Asymmetric snap-through time}
    \tilde{\tau}_\text{snap} = \frac{1}{\sqrt{\check{r}_\text{B}}}\ln\left(\frac{4}{\check{p}^A_0}\sqrt{\frac{2\check{r}_\text{B}}{\check{q}_\text{B}}}\right). 
\end{equation}
In terms of the original model variables,
the snap-through time for bifurcation buckling is
\begin{equation}\label{eq:tsnapA_original}
    t^\text{B}_\text{snap} = \frac{1}{\sqrt{\check{r}_\text{B}\DP}}\ln\left(\frac{4}{\check{p}^A_0\eta}\sqrt{\frac{2\check{r}_\text{B}}{\check{q}_\text{B}\DP}}\right),
\end{equation}
where a factor of $\DP^{\alpha}$ has been replaced with $\eta$. 
If we set $p_0^A = \check{p}_0^A\eta$ and we define $r_B=\lambda^{2}\check{r}_\text{B}$, and $q_B = \lambda^{2}\check{q}_\text{B}$, the snap-through time takes the form used in the main text as
\begin{equation}
    t^\text{B}_\text{snap} = \frac{\lambda}{\sqrt{r_B\DP}}\ln\left(\frac{4}{p^A_0}\sqrt{\frac{2 r_B}{q_B\DP}}\right),
    \label{eq:tsnap_A_main}
\end{equation}
where we make explicit the linear dependence of the snap-through time with $\lambda$, given that the quantities $r_B$ and $q_B$ tend to constants as $\lambda \to \infty$, as shown in Fig.~\ref{fig:rqB}.

\par
Although the expression in \eqref{eq:tsnapA_original} allows to compute the snap-through time for bifurcation buckling, it depends on the quantities $\check{r}_\text{B}$ and $\check{q}_\text{B}$, which are defined through infinite sums involving numerically computed eigenvalues and eigenfunctions. The numerical approximations involved in their computation can thus introduce significant errors. Here we show that the infinite sums for $\check{r}_\text{B}$ and $\check{q}_\text{B}$  can instead be expressed in terms of inner products of functions with closed-form analytical expressions.
We first introduce two functions $\vec{\varphi}_1$ and $\vec{\varphi}_2$ that are the solutions of the linear problems
\begin{subequations}\label{varphi eq}
\begin{align}
 \mathcal{L}\vec{\varphi}_1 &= \delta(\theta)\vec{F}, \\
 \mathcal{L}\vec{\varphi}_2 &= \mathcal{B}(\check{\vec{u}}_0^A,\check{\vec{u}}_0^A),
\end{align}
\end{subequations}
and for which an analytical form exists.  By writing those function via an eigenfunction expansions, we find that
\multig[phis]{
\vec{\varphi}_1(\theta) = \sum_{n\in\mathbb{S}}\frac{F_n}{\left(\omega_n^S\right)^2}\check{\boldsymbol{u}}_n^S,\\
\vec{\varphi}_2(\theta) = \sum_{n\in\mathbb{S}}\frac{\mathcal{B}_{00n}^{AAS}}{\left(\omega_n^S\right)^2}\check{\boldsymbol{u}}_n^S .
}
By comparing the forms \eqref{phis} with the ones in \eqref{asymmetric params}, the coefficients $\check{r}_\text{B}$ and $\check{q}_\text{B}$ can be rewritten in terms of integrals of analytical expressions as
\begin{align}\label{r_B ana}
    \check{r}_\text{B}= 2\sum_{n\in\mathbb{S}}\frac{F_n}{\left(\omega_n^S\right)^2}\mathcal{B}_{n00}^{SAA}  
    = 2\left\langle\mathcal{B}(\vec{\varphi}_1,\vec{\check{u}}_0^A),\vec{\check{u}}_0^A\right\rangle
\end{align}
and
\begin{align}
    \check{q}_\text{B} = \mathcal{T}_{0000}^{AAAA} + 2\sum_{n\in\mathbb{S}}\frac{\mathcal{B}^{SAA}_{n00}\mathcal{B}^{AAS}_{00n}}{\left(\omega_n^S\right)^2}
    =\mathcal{T}_{0000}^{AAAA} +2 \left\langle \mathcal{B}\left(\vec{\varphi}_2,\check{\vec{u}}_0^A\right),\check{\vec{u}}_0^A\right\rangle.
\end{align}
\begin{figure}
    \centering
    \includegraphics[width=\linewidth]{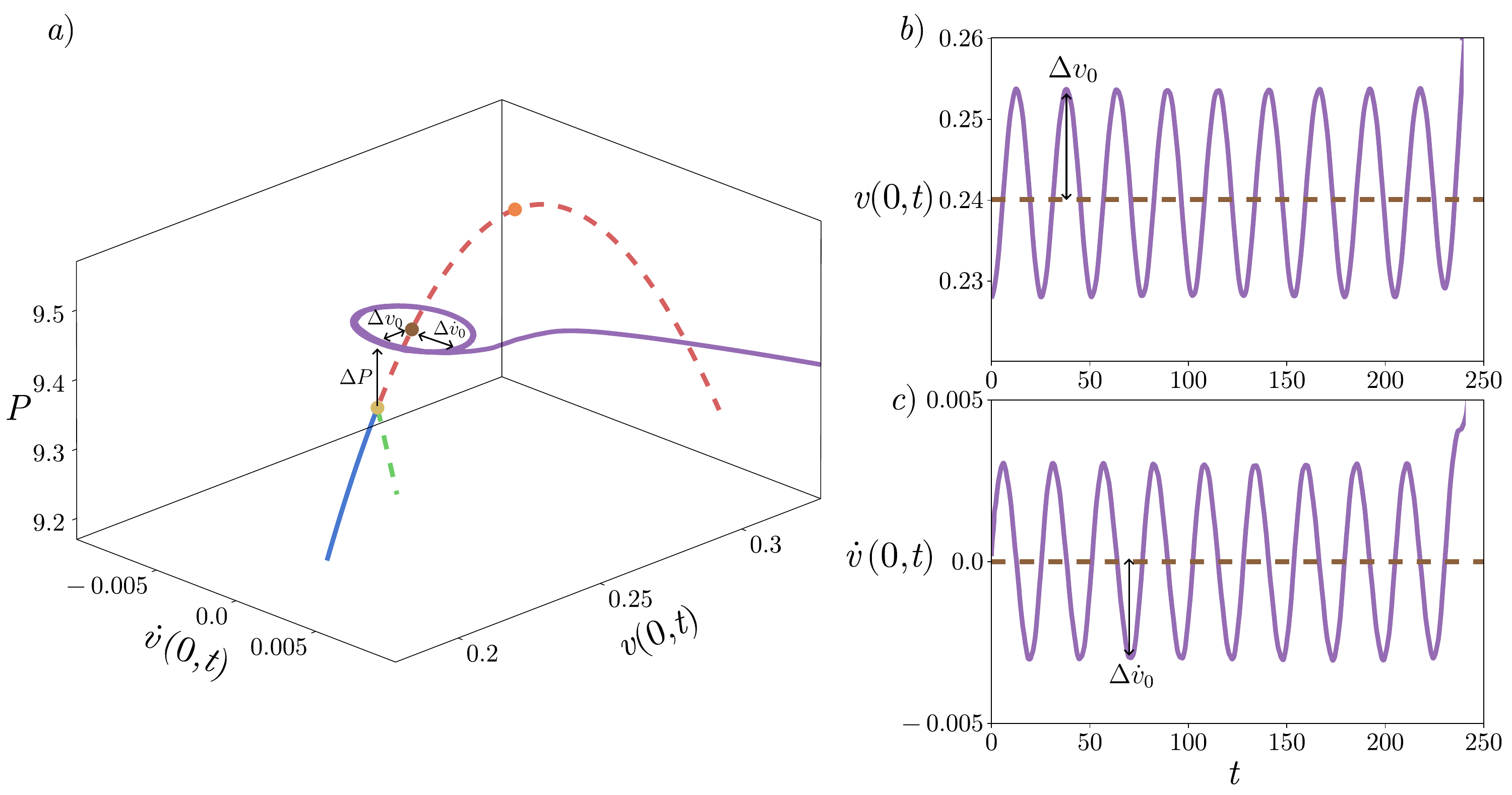}
    \caption{(a) Phase plane plot showing the trajectory of the mid-point of the arch after the applied load is increased by a quantity of $O(\DP)$. (b) Displacement of the mid-point of the arch against time. (c) Velocity of the mid-point of the arch against time.}
    \label{fig:Oscillation-plot}
\end{figure}
\section{Pre-snap-through oscillations}\label{Pre-Snap-Through Oscillations}

Numerical simulations of bifurcation buckling
reveal that there is a prolonged period where the arch
undergoes oscillatory behaviour before finally
snapping-through to the new stable equilibrium
configuration. Thus, a natural question to ask is: what configuration is the arch oscillating
about?  To answer this question, we use
numerical simulations to examine solution
trajectories in phase space when the applied load
is increased by an amount $\Delta P$ from the
critical value $P_{cr}$ associated with the
bifurcation buckling point.
Here, we are 
assuming that $P_{cr} + \Delta P$ remains below the
maximum load associated with the limit point
bifurcation of symmetric solutions.  
In Fig.~\ref{fig:Oscillation-plot}(a), we see that the system is moved to the symmetric unstable branch when the applied load is perturbed by an amount $O(\DP)$. The purple curve is a representative trajectory of the mid-point of the arch during the snap-through after the perturbation is applied. The oscillations, whose displacement $v$ and velocity $\dot{v}$ of the arch mid-point are shown in Fig. \ref{fig:Oscillation-plot}(b) and \ref{fig:Oscillation-plot}(c) respectively, represent the system orbiting around the globally unstable
symmetric configuration. Pre-snap-through oscillations are not seen in the limit-point buckling.

\begin{figure}
    \centering
    \includegraphics[width=1\linewidth]{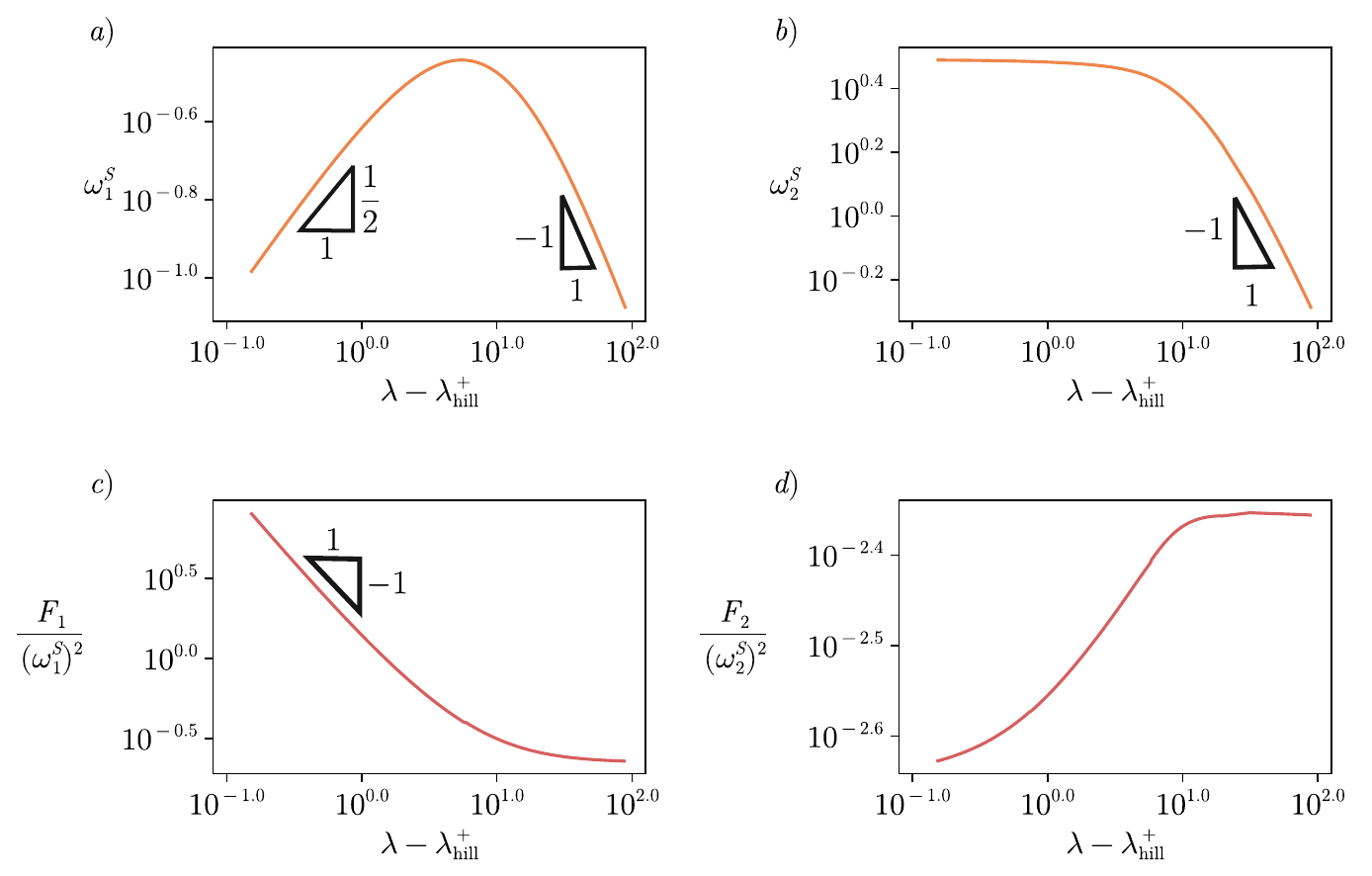}
    \caption{{Natural frequencies of the (a) first and (b) second symmetric mode and amplitudes of the (c) first and (d)
   second symmetric mode as a function of the distance from the hill-top bifurcation $\lambda-\lambda_{hill}^+$. Triangles show power law relations; for  $\lambda$ large, we recover the $\omega_n^{S} = O(\lambda^{-1})$  dynamics as expected from the scaling arguments presented in the main text.}}
    \label{fig:oscillation consts}
\end{figure}

{From the multiple-scales analysis of bifurcation buckling, we find that the time-dependent expression for the symmetric modes given in eq.~\eqref{A01 solution perturbed oscillation} describes the pre-snap-through oscillations. These oscillations are characterized by amplitudes $F_n / (\omega_n^S)^2$ and natural frequencies $\omega_n^S$.  The
natural frequency increases as the mode number $n$ increases and hence the oscillation
amplitude decreases: we show this behaviour by comparing the natural frequencies of the first and second modes in Fig.~\ref{fig:oscillation consts}~(a)--(b) and their amplitudes in Fig.~\ref{fig:oscillation consts}~(c)--(d). Hence,
despite the central point load exciting an infinite number of
oscillatory symmetric modes, only the first mode ($n = 1$) is readily seen.}
% In Fig.~\ref{fig:oscillation consts}, we plot the values of $\omega_{1,2}^S$, and the prefactors $F_{1,2}/\left(\omega_{1,2}^S\right)^2$, this shows why, despite an infinite spectrum of oscillations, only the first mode is readily seen, since the amplitudes decrease sharply after the first symmetric mode. 

{As the hill-top bifurcation point is approached, 
$\lambda\to\lambda_{\text{hill}}^+$, the natural frequency
of the first oscillatory mode, $\omega_1^{S}$, tends to zero with
$\omega_1^{S} = O\left((\lambda - \lambda_{\text{hill}}^{+})^{1/2}\right)$.  When $\lambda = \lambda_{\text{hill}}^{+}$, the
linearised system has a double zero eigenvalue, as both the first symmetric and antisymmetric eigenvalues are zero.  
The increase in oscillation period that occurs as the hill-top bifurcation
is approached is analogous to the increase in snap-through
time as a limit-point or pitchfork bifurcation is approached and hence
represents a new type of critical slowing down.  As $\lambda$ is increased
from $\lambda_\text{hill}^{+}$, the natural frequency $\omega_1^{S}$ 
increases until a maximum is reached, after which it decreases.  
% \emph{The maximum in frequency represents the transition between
% stretching- and bending-dominated dynamics.} {
The natural frequency $\omega_2^{S}$, instead, remains constant before decreasing.  For large $\lambda$, the natural frequencies scale like $\omega_n^{S} = O(\lambda^{-1})$, consistent with the scaling analysis presented in the main text when balancing inertia and bending contributions. The behaviour that we see in the frequency $\omega_1^{S}$ can be explained by a combination of effects resulting from the hill-top bifurcation and the transition between stretching- and bending-dominated dynamics.}

% When $\lambda = \lambda_{\text{hill}}^{+}$, the
% linearised system has a double zero eigenvalue, as both the first symmetric and antisymmetric eigenvalues are zero.  A double zero eigenvalue is
% consistent with
% a co-dimension 2 bifucation point, which occurs when the limit-point and
% pitchfork bifurcations coincide. 

{Due to our assumption that the
imperfection introduced into the system is $O(\Delta P)$ in size, 
the imperfection does not affect the linear operator $\mathcal{L}$.
Hence, the amplitudes and natural frequencies of the pre-snap-through
oscillations are not influenced by the imperfection.}

{\section{Overshooting the limit-point bifurcation}}\label{Sec:Overshoot}
{Numerical simulations reveal there is a small range of slenderness values $\lambda$ exceeding $\lambda_{\text{hill}}^{+}$ where the arch evolves symmetrically during snap-through and the snap-through time 
is much smaller than that predicted for bifurcation buckling (see Fig.~1~(e) in the main text).
The dynamics in this region resemble snap-through associated with a limit-point bifurcation.}

\begin{figure}
    \centering
    \includegraphics[width=\linewidth]{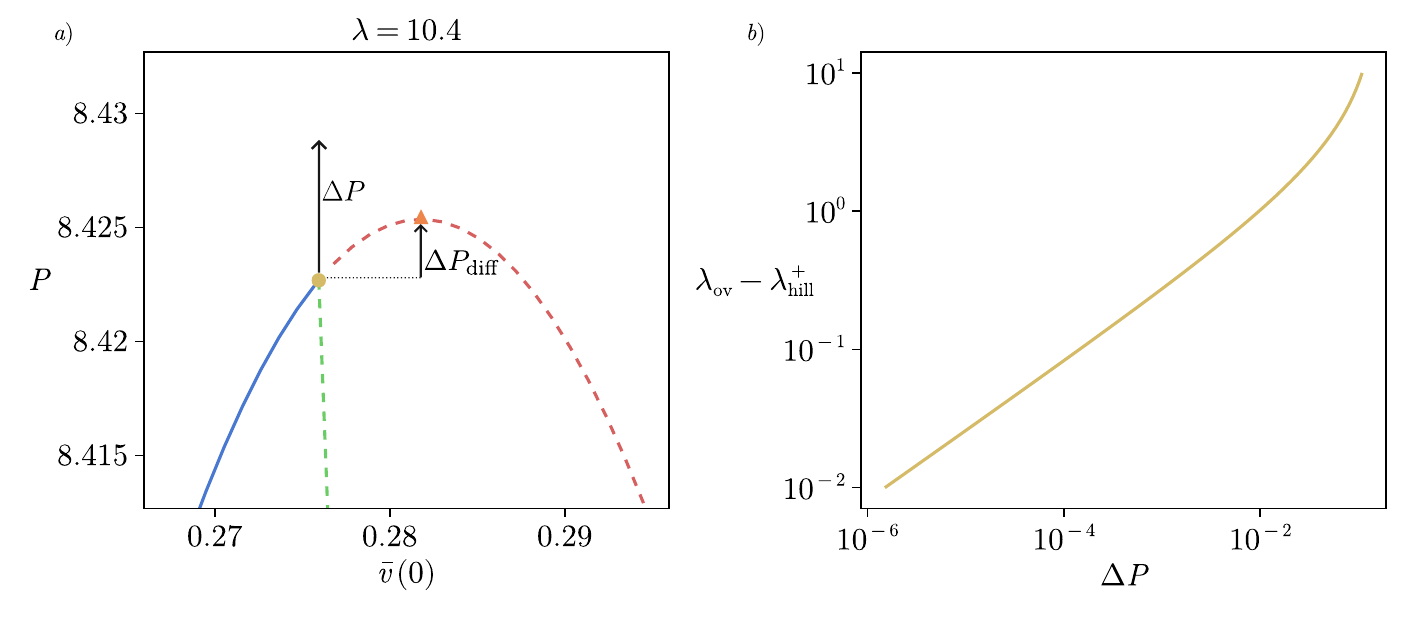}
    \caption{{Overshooting the limit-point
    bifurcation.  (a) Force-displacement curve obtained with $\lambda = 10.4$ where the critical loads for bifurcation buckling and limit-point buckling are denoted by a circle and triangle, respectively, and their difference denoted by $\DP_{\text{diff}}$. Applying a load increment of size $\Delta P$ to the critical load for bifurcation buckling can lead to an
    overshoot of the limit-point bifurcation when $\Delta P > \DP_{\text{diff}}$. (b) Width of the overshoot region, $\lambda_\text{ov} - \lambda_{\text{hill}}^{+}$, as a function of the load increment $\Delta P$.}}
    \label{fig:Zoom}
\end{figure}

{For slenderness values
$\lambda > \lambda_{\text{hill}}^{+}$, the system exhibits two critical loads: one
associated with bifurcation buckling and one associated with limit-point buckling. 
When $\lambda$ slightly exceeds $\lambda_{\text{hill}}^{+}$, the critical load
for bifurcation buckling is only slightly less than the critical load for limit-point buckling.
Therefore, when the applied load is increased beyond the threshold for bifurcation buckling by an amount $\Delta P$,
the threshold for limit-point buckling can also be exceeded as depicted in Fig.~\ref{fig:Zoom}~(a).  We call this
`overshooting' the limit-point bifurcation.  When overshooting
occurs, the response of the system becomes dominated by the
limit-point bifurcation rather than the pitchfork bifurcation.
Hence, the arch remains symmetric during snap-through and the snap-through
time is reduced.}

{To determine when overshoot occurs, we fix the
arch slenderness to $\lambda > \lambda_{\text{hill}}^{+}$ so that
both bifurcation and limit-point buckling occur.  We then calculate
the difference in the critical loads of the two buckling modes,
$\Delta P_{\text{diff}}$, as illustrated in Fig.~\ref{fig:Zoom}~(a).
A load increment of size $\Delta P$ applied to the critical configuration
associated with bifurcation buckling will cause an overshoot
if $\Delta P > \Delta P_{\text{diff}}$. 
The value of $\Delta P_\text{diff}$ will be a function of $\lambda$.  
By inverting this function, it is possible to determine the maximum
value of the arch slenderness, $\lambda_{\text{ov}}$, that will lead to overshoot if a 
load increment of size $\Delta P$ is applied to the system.  
Thus, for a fixed load increment $\Delta P$, the region of overshoot is 
then given by $\lambda_\text{hill}^{+} < \lambda < \lambda_{\text{ov}}(\Delta P)$.  The width of the overshoot region $\lambda_{\text{ov}} - 
\lambda_{\text{hill}}^{+}$ initially grows like ($\Delta P)^{1/2}$,
as seen in Fig.~\ref{fig:Zoom}~(b).  As the overshoot region is determined
solely through a consideration of the equilibrium configurations of the arch,
it is not impacted by the imperfections we introduce into the system.
An extended multiple-scales analysis in 
the overshoot region could be carried out to calculate the snap-through
time; however, we leave this as an area of future work.}

\section{Alternative types of imperfections}
\label{sec:geo_imperfections}

The analytical framework developed here can be applied to mechanical systems with imperfections in the geometry, rather than the
initial conditions.  For example, bifurcation buckling of shells is known to be sensitive to geometric imperfections~\cite{ShellImperfections1,ShellImperfections2}.  
A geometric imperfection with non-dimensional size $\epsilon \ll 1$ will
change the critical load at which buckling occurs by an amount
proportional to $\epsilon^{2/3}$~\cite{koiter1970}.
% In particular, if the geometric imperfection has non-dimensional
% size $\epsilon\ll1$, then the critical load at which buckling occurs
% changes by an amount that is proportional to $\epsilon^{2/3}$.
For a geometrically imperfect arch subjected to a central point load~\cite{THOMPSON1983,Hunt1977,Thompson2017}, the critical load 
can be written as
$P_{cr}  = P_{cr}^{(0)}+\epsilon^{2/3} P_{cr}^{(2/3)}+O\LR{\epsilon^{4/3}}$, 
where $P_{cr}^{(0)}$ is the critical load for the perfect system. Using a dominant balance argument, the critical configuration at buckling for the imperfect arch can be written as
\begin{equation}\label{eq:ImpSensitivity}
\vec{u}_{cr}(\theta) = \vec{u}_{cr}^{(0)}(\theta)+\epsilon^{1/3}\vec{u}_{cr}^{(1/3)}(\theta)+O\LR{\epsilon^{2/3}},
\end{equation}
where $\vec{u}_{cr}^{(0)}$ is the critical configuration for the perfect arch calculated in Sec.~\ref{sec:equilibrium}.

The dynamics of bifurcation buckling can be captured using the
multiple-scales analysis developed in Sec.~\ref{sec:BifBuck}
by transferring the imperfection introduced statically in the geometry to an equivalent imperfection in the initial conditions. To show this equivalence,
we first write the displacement field in a similar way to eq.~\eqref{eq:expansions}, namely
\begin{equation}\label{eq:ExpansionImperfect}
\boldsymbol{u}(\theta,t) = \boldsymbol{u}_{cr}(\theta)+\DP\breve{\vec{u}}(\theta,t),
\end{equation}
where $\DP = (P_{cr} - P) / P_{cr} \ll 1$.  However, contrary to
\eqref{eq:expansions}, the quantities $\vec{u}_{cr}$ and $P_{cr}$ in \eqref{eq:ExpansionImperfect} now correspond to the critical configuration
and buckling load for the imperfect arch.  
The dynamical problem associated with $\breve{\vec{u}}$
satisfies homogeneous initial conditions because the imperfection is now assumed to enter through the geometry.  We now write the solution
exactly as in \eqref{eq:expansions}, i.e.\ as $\vec{u}(\theta,t) = \vec{u}_{cr}^{(0)}(\theta) + \DP \hat{\vec{u}}(\theta,t)$,
by substituting the expansion in eq.~\eqref{eq:ImpSensitivity} into \eqref{eq:ExpansionImperfect} 
% and recalling that, in the approach used in this work, the displacement field is written as $\vec{u}(\theta,t) = \vec{u}_{cr}^{(0)}(\theta) + \DP \hat{\vec{u}}(\theta,t)$, we find that $\hat{\vec{u}}$ 
% can be defined as
% becomes, in terms of the geometrical imperfection,
and then defining $\hat{\vec{u}}$ as 
\begin{equation}
\hat{\vec{u}} \equiv \breve{\vec{u}} + \DP^{-1} \epsilon^{1/3} \vec{u}_{cr}^{(1/3)} + O(\DP^{-1}\epsilon^{2/3}). 
\end{equation}
Since $\breve{\vec{u}}$ satisfies homogeneous
initial conditions, $\hat{\vec{u}}$ satisfies inhomogeneous initial conditions given by
\begin{equation}
    \hat{\vec{u}}(\theta,0) = \DP^{-1} \epsilon^{1/3} \vec{u}_{cr}^{(1/3)}(\theta) + O(\DP^{-1}\epsilon^{2/3}), 
    \qquad
    \pdv{\hat{\vec{u}}}{t}(\theta,0)=0.
\end{equation}
Setting $\epsilon = \DP^{3} \eta^3$ with $\eta = O(\DP^{\alpha})$
and $\alpha \geq 1$ then recovers the imperfect
initial conditions used in eq.~\eqref{eq:icimpefection_hat}.  
The multiple-scales analysis of $\hat{\vec{u}}$ will
proceed exactly as in Sec.~\ref{sec:BifBuck}.  
The snap-through time will be given by
eq.~\eqref{eq:tsnapA_original} with an imperfection parameter defined as
$\check{p}_0^A = \langle \vec{u}_{cr}^{(1/3)}, \check{\vec{u}}_0^A \rangle$.
Thus, it is only the perturbation to the critical buckling configuration
that sets the snap-through time.  It can be shown that perturbation to the critical load induced by the imperfection does not impact the snap-through time as it enters the problem at a higher order.

Although we focus on the behaviour of an arch, our analysis can be extended to snap-through in shells and other two-dimensional objects~\cite{huang2024}. The extra spatial dimension will likely enrich the bifurcation landscape of
shells.  Nevertheless, shells exhibit both bifurcation buckling and imperfection sensitivity, thus making the physics of the arch that are revealed here applicable to their analysis.

\bibliographystyle{unsrt}
\bibliography{sample_arxiv}

\end{document}